\documentclass[preprint]{elsart3p}

\usepackage{dcolumn}% Align table columns on decimal point
\usepackage{bm}% bold math
\usepackage{graphicx}        % standard LaTeX graphics tool when including figure files
\usepackage{cite}            % adjusts the "syntax" of the refs in the text
\usepackage{bm}
\usepackage{amsmath}
\usepackage{amssymb}

\usepackage{amssymb}
\journal{Journal of Magnetic Resonance}

\begin{document}

\begin{frontmatter}
\date{}


\title{Numerical simulations of NMR relaxation in chalk \\ using local Robin boundary conditions} 


\author{M. \"{O}gren$^{a,b,*}$, D. Jha$^{a}$, S. Dobbersch\"{u}tz$^{a}$, D. M\"{u}ter$^{a}$,}
\author{M. Carlsson$^{c}$, M. Gulliksson$^{b}$, S. L. S. Stipp$^{a}$, and H. O. S\o rensen$^{a}$}

\address{$^{a}$Nano-Science Center, Department of Chemistry, University of
Copenhagen, Universitetsparken 5, 2100 K\o benhavn {\O}, Denmark. \\
$^{b}$School of Science and Technology, \"{O}rebro University,
701 82 \"{O}rebro, Sweden. \\
$^{c}$Center for Mathematical Sciences, Lund University, Box 118,
22100 Lund, Sweden.\\
$^{*}$Corresponding author: magnus@ogren.se}

\begin{abstract}
The interpretation of nuclear magnetic resonance (NMR) data is of interest in a number of fields. In \"{O}gren {[}Eur. Phys. J. B (2014) \textbf{87}: 255{]} local boundary conditions for random walk simulations of NMR relaxation in digital domains were presented. Here, we have applied those boundary conditions to large, three-dimensional (3D) porous media samples. We compared the random walk results with known solutions and then applied them to highly structured 3D domains, from images derived using synchrotron radiation CT scanning of North Sea chalk samples. As expected, there were systematic errors caused by digitalization of the pore surfaces so we quantified those errors, and by using linear local boundary conditions, we were able to significantly improve the output. We also present a technique for treating numerical data prior to input into the ESPRIT algorithm for retrieving Laplace components of time series from NMR data (commonly called $T$-inversion).
\end{abstract}


\end{frontmatter}

% \linenumbers


\section{Introduction}

Simulation of particle diffusion through complex media can be used
to model many physio-chemical processes in biology, groundwater transport,
filtration and for many applications in industry~\cite{mueter2015}.
One can also use the simulated movement of particles to represent
diffusing molecules, heat transport, or as here, nuclear magnetic
resonance (NMR) excitations carried by protons. NMR based research
is of interest in many fields, for example to depict brain tumors~\cite{LinAPL2013},
in biofilm purification of contaminated water~\cite{Fridjonsson2011},
in manufactured porous fiber materials~\cite{Tomadakis2003}, as
well as in petrophysical applications~\cite{Senturia1970,Cohen1982,Kenyon1992}.

When particles interact with a surface or other particles, reactions,
reflections or sorption can happen, which changes or interupts the
movement of the particles. To describe such systems, a variety of
probabilistic mathematical models are used. In the literature, there
are reports of the interplay between theoretical and experimental
work made on ideal systems of simple geometry, such as glass triangles~\cite{Finjord2007}
(and references therein), where exact solutions can be obtained with
geometry specific methods. There are also large scale numerical simulations
of (digitized) true porous media that have been made by practitioners
in various fields, where predictions have been compared with results
from experiments. One example is data from core plugs of porous rocks~\cite{Vincent2011,Mohnke2014}.
From what is reported in the literature, quantitative agreement between
simulations and measurements is often obtained only after fitting
the two data sets with a free parameter or after optimising the surface
relaxation parameter, $\rho$, with the experimental data~\cite{OerenSPE2002,Talabi2009}.
The value of the surface relaxation parameter used in models strongly
affects the NMR decay rate and subsequently the estimated properties
of the medium. In reality surface relaxivity is difficult to measure.
Therefore most NMR correlations assume constant $\rho$, even though
it is straightforward in stochastic particle models to allow it to
be space dependent. For heterogeneous rocks composed of more than
one mineral, $\rho$ is not constant and surface relaxivity is reported
to increase with higher fractions of microporosity~\cite{Kenyon1992}.

Despite of how we model the relaxivity, systematic errors may arise
in calculated physical properties, because independently of how accurate
we solve the problem for the digital domain, the true domain can be
quite different from its digitalization. In applications of NMR on
rock samples, relaxation is generally slow in the bulk pore volume,
$V$, and relatively faster at the pore surface. Therefore, the description
of the surface geometry is essential for simulations. However, in
digital domains the surface area, $S$, does not converge to the true
value when resolution increases, in contrast with the behaviour of
pore volume, and the methods for treating the surfaces in numerical
simulations are critical. Different approaches have been suggested
to meet this difficulty~\cite{BergmannPRE1995,JinJoMR2009}. 

This study extends our previous work~\cite{Ogren2014}, that used
local boundary conditions (LBC) to locally adapt the probability for
surface relaxation, $p_{S}$, when a NMR excitation encounters the
pore surface in the porous medium. For simple domains we have shown
that this method closely reproduces the solutions to the corresponding
PDE model for which the surface relaxation is controlled by the surface
relaxation parameter $\rho$ via a Robin boundary condition~\cite{Ogren2014}.
It is clear that development towards more accurate numerical modeling
benefit from predictions on well characterised samples, and the purpose
of this article is to first carry out benchmarking simulations and
then produce results for natural chalk samples. 

By starting from two-dimensional artificial porous media of moderate
size, we can directly evaluate LBC against results obtained using
the finite element method (FEM). For three dimensions we then benchmark
the method against analytic solutions for two simple geometries, the
ball and the cube. We also rotate the cube with respect to the coordinate
system of the digital domain to investigate how LBC works when the
surfaces of the cube is not aligned with the faces of the voxels.
Finally, the algorithm is used to calculate the NMR relaxation in
two different complex geometries of chalk imaged by X-ray tomography
and consisting of approximately $10^{3}\times10^{3}\times10^{3}=10^{9}$
voxels.

\subsection{NMR relaxation model\label{sub:NMR-relaxation-model}}

The fundamentals behind nuclear magnetic resonance rely on the quantum
mechanical magnetic properties of specific atomic nuclei~\cite{Rabi1938},
hence \textquotedblleft{}nuclear\textquotedblright{} and \textquotedblleft{}magnetic\textquotedblright{}.
When such a magnet (being placed in an external magnetic field) absorbs
energy, the nucleus is described as being in \textquotedblleft{}resonance\textquotedblright{}.
This apply, for example, to a porous media filled with water, where
the hydrogen nucleus have an intrinsic magnetic moment that can be
excited with a radio frequency electromagnetic pulse. The process
when the nuclei return to the non-excited state is called relaxation.
NMR relaxation analysis of porous media can be used, for example,
to estimate permeability, the pore size distribution, and identify
pore fluids and gases.

The so called Bloch-Torrey equations~\cite{BlochPhysRev1946,TorreyPhysRev1956}
phenomenologically describe the relaxation dynamics of excited nuclear
magnetic moments in the general case. In the case of isotropic diffusion
and in the absence of external magnetic field gradients, we get after
the termination of the magnetic excitation the remaining single partial
differential equation (PDE) describing a magnetic moment $M\left(\mathbf{x},t\right)$~\cite{Senturia1970,Cohen1982,BrownsteinPRA1979,MohnkeVZJ2010}
in a pore domain $\Omega$

\begin{equation}
\frac{\partial M}{\partial t}=D_{0}\nabla^{2}M-\frac{M}{T_{V}}.\label{eq:Diffusion_equation}
\end{equation}
This is formally a time dependent heat equation with a source term
that is proportional to the magnetic moment. In (\ref{eq:Diffusion_equation})
$D_{0}$ is the diffusion constant and $T_{V}$ is the characteristic
time of relaxation in the volume, often called bulk relaxation. At
the boundary $\partial\Omega$, we have a mixed Robin boundary condition
(BC)

\begin{equation}
D_{0}\mathbf{n}\cdot\nabla M+\rho M=0,\label{eq:Robin}
\end{equation}
where $\rho$ is the surface relaxation parameter. Finally, we need
to define an initial condition for the magnetic moment, which we choose
to be constant $M\left(\mathbf{x},0\right)=M_{0}$ throughout this
article. In e.g.~\cite{Finjord2007,Ogren2014} other initial conditions
are investigated. Uniform initial magnetisation is common in experiments
but nonuniform initial conditions can be designed using inhomogeneous
radio frequency pulses~\cite{SongPRL2000}. In the above PDE model
we have assumed that $T_{V}=T_{V,2}$ ($T_{V}=T_{V,1}$) and $\rho=\rho_{2}$
($\rho=\rho_{1}$) are for the transverse (longitudinal) component
of nuclear spin magnetisation~\cite{Kenyon1992,Grebenkov2007}, i.e.,
we consider tranverse spin components in the following. Realistic
values for the above physical parameters in the three-dimensional
case were taken from the literature~\cite{Talabi2009} (and references
therein), $D_{0}\simeq2.1\cdot10^{-9}$~m$^{2}$/s, $T_{V}\simeq3.1$~s
and $\rho\simeq1.0\cdot10^{-5}$~m/s. The first two parameters refer
to water (brine) and the last to a representative brine-chalk interface~\cite{Vincent2011,Mohnke2014}.
For NMR experiments of materials chemically sensitive to water, such
as foods, polymers and aerogels, gases (e.g., C$_{2}$F$_{6}$ and
$^{129}$Xe) can alternatively be used~\cite{LizakJMR1991,SongJMR1995}.

In addition to $T_{V}$, there are two other characteristic times
in the problem. By dimensional analysis we find the diffusion time,

\begin{equation}
T_{D_{0}}\sim R_{0}^{2}/D_{0},\label{eq:Diffusion-time}
\end{equation}
and the surface relaxation time,

\begin{equation}
T_{\rho}\sim R_{0}/\rho,\label{eq:Surface-relaxation-time}
\end{equation}
which suggest two limiting regimes dependent on how the typical pore
radius $R_{0}$ relate to the other parameters. We have \emph{fast
diffusion} when $T_{D_{0}}\ll T_{\rho}$, i.e., with the parameters
above, approximately for $R_{0}\ll100$ $\mu$m, and \emph{slow
diffusion} in the opposite limit where $T_{D_{0}}\gg T_{\rho}$ and
$R_{0}\gg100$ $\mu$m. Qualitatively, in the \emph{fast diffusion
}regime particles carrying a magnetic moment are diffusing to surfaces
faster than they are annihilated, resulting in a flat distribution
of $M\left(\mathbf{x},t\right)$ for all times, while with \emph{slow
diffusion} the number of particles is less close to surfaces and $M\left(\mathbf{x},t\right)$
develops large variations when $\mathbf{x}$ is in the vicinity of
the surface. Therefore, if the typical pore sizes are known in respect
to the physical parameter values, one can qualitatively predict the
dynamics.

Experimentally one often measures the total magnetisation, obtained
by integration of the magnetic moment per unit volume\textbf{ }over
the pore domain $\Omega$

\begin{equation}
\mathcal{M}\left(t\right)=\int_{\Omega}M\left(\mathbf{x},t\right)d\mathbf{x}.\label{eq:Def_of_total_magnetization}
\end{equation}
In this article we study the time dependence of~(\ref{eq:Def_of_total_magnetization})
with various methods and boundary conditions.

The article is briefly outlined as follows. In Sect. \ref{sec:Method}
we present the method we have used to solve the PDE model for NMR
relaxation presented above. We tested the numerical method against
known results in Sect.~\ref{sec:Tests-of-the-new-numerical-method}
and in Sect.~\ref{sec:Application-to-NMR-relaxation-in-true-porous-media}
we apply the method to NMR relaxation in natural chalk samples and
analyse the data in terms of real exponential components. Finally,
we present a discussion and a summary of the results in Sect.~\ref{sec:Discussion-and-summary}.

\section{Method\label{sec:Method}}

As pioneered by Kolmogorov, Feynman, Kac, and many others, one can
define stochastic processes for many second order PDEs that in the
limit of many random realisations may converge to either stationary
or time dependent solutions of the PDE~\cite{KaratzasBook1991,RiskenBook1996,KampenBook2008,GardinerBook2009}.
A well known example is Eq.~(\ref{eq:Diffusion_equation}).
In fact the derivation of a diffusion equation is often motivated
by a random walk of quasi-particles given by a chemical concentration,
a quantity of heat, or as here a magnetic moment carried by protons
in water molecules. The additional last term, $-M/T_{V}$ in~(\ref{eq:Diffusion_equation}),
is simply modelled using the probability $p_{V}=\Delta t/T_{V}$ for
the annihilation of every random walker at each time step $\Delta t$.
However, as long as $T_{V}$ is considered to be a time and space
independent constant, the effect of volume relaxation ($V$) can be
factorized out from the dynamics of the total magnetisation, such
that
$$
M_{T_{V}}\left(\mathbf{x},t\right)=\mathrm{e}^{-t/T_{V}}M_{\infty}\left(\mathbf{x},t\right),
$$
\begin{equation}
M_{T_{V}}\left(\mathbf{x},t\right)\rightarrow M_{\infty}\left(\mathbf{x},t\right),\: T_{V}\rightarrow\infty\label{eq:T_V_factorisation}
\end{equation}
holds for any solution of~(\ref{eq:Diffusion_equation}). Therefore,
hereafter we do not consider the effect of $T_{V}$. In the computer
implementation this is done by choosing the volume relaxation time
very large ($T_{V}\rightarrow\infty$). Consequently, for the results
for the chalk samples presented later in the article (Fig.~\ref{fig:real_samples})
it is necessary to multiply the total magnetisation with a factor
$\mathrm{e}^{-t/T_{V}}$ in order to compare with experimental measurements.

\subsection{Cartesian random walk}

The idea here is to use a linear LBC (LLBC) relation between the parameter
$\rho$ and the corresponding parameter $p_{S}$ in a Cartesian lattice
based random walk implementation of the problem for a boundary with
arbitrary curvature. There are in fact infinitely many degrees of freedom in formulating a corresponding stochastic
process for the PDE~(\ref{eq:Diffusion_equation}). For example different
distributions of random numbers can be used, and so called gauges
can give varying numerical properties while converging to the same
solution~\cite{DrummondICOLS2003,OgrenCPC2011}. Here we use a Cartesian
random walk with isotropic step lengths $\Delta r\equiv\Delta x=\Delta y=\Delta z$
where $\Delta r$ is the resolution (voxel size) of the digital image
(Fig.~\ref{fig:3D_CT-images}). 
While the computational effort can
be dramatically reduced by the introduction of a distance operator
to take larger steps when the walkers are far from boundaries~\cite{TorquatoAPL1989}
we do not exercise this technique here but focus on illustrating the
effect of the implemented local boundary conditions. Moreover, the
uniform initial condition $M\left(\mathbf{x},0\right)=M_{0}$ means
that each voxel in the pore-domain $\Omega$ is equally likely to
be occupied by a random walker at $t=0$. For a walker at position
$\left(x,y,z\right)$ and time $t\geq0$ we then have for $t=t+\Delta t$
that 
\begin{equation}
\left(x,y,z\right)=\left(x,y,z\right)+\Delta r\vec{{r}}_{j},\label{eq:Definition_of_a_step_I}
\end{equation}
where 
\begin{equation}
\Delta t=\frac{\left(\Delta r\right)^{2}}{6D_{0}}.\label{eq:Definition_of_a_step}
\end{equation}
In equation (\ref{eq:Definition_of_a_step_I}) $\vec{{r}}_{j},\: j=1,2,...,6$
is one of the six unit vectors $\pm\vec{{e}}_{j},\: j=1,2,3$ and
$p(\vec{{r}}_{j})=1/6$, or in other words the probability of taking
a step of length $1$ or $-1$ in either of the three Cartesian directions
is equal. 

The total magnetisation (\ref{eq:Def_of_total_magnetization}) in
each time step is equal to the number of active trajectories, i.e.,
the sum over the remaining walkers. This number is normalized in each
timestep with the initial number of walkers. Hence the only information
needed is the time when a walker is annihilated, which can be easily
implemented for parallel computation. Consider a walker taking a step
according to~(\ref{eq:Definition_of_a_step}) that will cross a boundary.
Then it has to be decided if the walker is going to be annihilated
or reflected back to the original position. In the PDE model for NMR
relaxation it is the value of the surface relaxation parameter $\rho$
in Eq.~(\ref{eq:Robin}) that determines the effect of surface
relaxation. More precisely, the BC, Eq.~(\ref{eq:Robin}), is
Dirichlet if $\rho\rightarrow\infty$ and Neumann if $\rho\rightarrow0$.
For any other value, $0<\rho<\infty$, we need to relate the surface
relaxation parameter in the PDE model to a probability $p_{S}$ for
annihilation of a walker crossing a boundary surface ($S$). The linear
relation 
\begin{equation}
p_{S}=\frac{\Delta r}{D_{0}}\rho,\label{eq:p_S-rho-relation}
\end{equation}
and other relations, are commonly discussed~\cite{Finjord2007,BanavarPRL1987,MendelsonPRB1990,BergmannPRE1995}.
A derivation of Eq.~(\ref{eq:p_S-rho-relation}) can be found
in Sect.~3 of~\cite{Ogren2014}.

\subsection{Local Robin boundary conditions for digital domains\label{sub:Local-Robin-boundary-conditions}}

We first introduce the three-dimensional digital pore-domain $\Omega$
in which the diffusion-relaxation dynamics are considered, via the
binary phase function 
\begin{equation}
Z\left(\mathbf{x}\right)=\left\{ \begin{array}{l}
1,\,\,\mathbf{x}\in\Omega\\
0,\,\,\textnormal{otherwise}
\end{array}\right.,\label{eq:Def_Pore_Domain}
\end{equation}
where $\mathbf{x}=(x,y,z)$. 
Characterizations of the digitized media,
such as porosity can then be calculated from~(\ref{eq:Def_Pore_Domain})
as $\phi=\sum_{k=1}^{N^{3}}Z\left(\mathbf{x}_{k}\right)/N^{3}$, where
$\mathbf{x}_{k}$ is the coordinate for the center of voxel number
$k$ and $N\Delta r$ is the side length of the cubic media, see Table~\ref{Table_for_real_samples}
for examples. From Eq.~(\ref{eq:Def_Pore_Domain}) we define
the boundary $\partial\Omega$ as the two-dimensional pore-surface
with one of the outgoing unit vectors $\pm\vec{{e}}_{j},\: j=1,2,3$
situated a distance $\Delta r/2$ outside the pore-domain, i.e. for
which

\begin{equation}
Z\left(\mathbf{x}\right)=1,\,\, Z\left(\mathbf{x}\pm\Delta r\vec{{e}}_{j}\right)=0.\label{eq:Def_Pore_Surface}
\end{equation}
Therefore, the digital surface of the cubic media can be calculated
as the point set that fulfills Eq.~(\ref{eq:Def_Pore_Surface}).

In order to construct the linear local boundary conditions (LLBC)
for a point $\mathbf{x}$ fulfilling Eq.~(\ref{eq:Def_Pore_Surface})
we need to distinguish between the $2^{8}=256$ possible lattice configurations
for each cell surrounding $\mathbf{x}$ in each diagonal direction
that is in contact with a boundary~\cite{Ogren2014}. 
For this purpose
we can locally define the integer $I\left(\mathbf{x}\right)=\sum_{j=1}^{8}Z\left(\mathbf{x}_{j}\right)2^{8-j}\in\left\{ 0,1,...,255\right\} $,
where $\mathbf{x}_{j}$ surrounds $\mathbf{x}$. For the $256$ configurations
we are locally going to interpolate the corners of the digital media
by constructing new surfaces build up by triangular shapes. 
The procedure
we outline here for locally generating these improved linear boundaries
is equivalent to Marching cubes~\cite{Newman2006}. However, in our
case we have two choices. We can calculate all such surfaces in the
domain, store them in a lookup table, and use their reduction factor
in surface area together with the boundary condition of Eq.~(\ref{eq:p_S-rho-relation}).
Alternatively, we can let each walker that first interacts with a
given boundary perform this procedure on the fly, which is clearly
more efficient when the number of walkers are few.

\subsection{Outer boundaries}

The domains of the chalk samples may be large but are always finite.
Hence we need to treat walkers that escape the cubic ($N^{3}$) domain
via pore-voxels in one of the six outer surfaces. 
The method we use
is to introduce a walker in a random pore-voxel in the opposite outer
surface of which it escaped. This procedure is common in the literature~\cite{OerenSPE2002,Talabi2009}
and our simulation results are in fact fairly insensitive to the way
we treat those walkers escaping the domains. 
This is a consequence
of the large cubic domains we use and therefore the probability of
escaping the domain is much smaller than to interact with the pore
surface.

\subsection{Statistical accuracy of the solution}

When discussing the convergence of a stochastic simulation as outlined
above we generally need to fulfill the following conditions. 
\begin{description}
\item [{(1)}] The number of trajectories should be sufficient to obtain
results of a certain statistical significance. 
\item [{(2)}] The step size should be small enough in order to probe all
of the small scale geometry of the media. Note that this is directly
connected to the resolution used when producing the digital media. 
\end{description}
If conditions (1) and (2) are fulfilled, one can accurately simulate
diffusion processes with Dirichlet ($\rho\rightarrow\infty$) and
Neumann ($\rho\rightarrow0$) boundary conditions in Eq.~(\ref{eq:Robin}).
In the context of NMR relaxation this limits would mean that all particles
were affected by surface relaxation ($\rho\rightarrow\infty$), or
that all were reflected back uneffected ($\rho\rightarrow0$).

However, for Robin boundary conditions ($0<\rho<\infty$) we require
the following additional probability based modeling of the surface
relaxation. 
\begin{description}
\item [{(3a)}] A relation (algorithm dependent) between the local probability
for surface relaxation, $p_{S}$, and the function $\rho\left(\mathbf{x},t\right)$
describing the local surface relaxation parameter. 
\item [{(3b)}] A local description of the surface area for a digital media. 
\end{description}
We believe that neither (3a) or (3b) is always satisfactorily treated
in the literature. Although (3a) has been discussed for some specific
algorithms and geometries, e.g.,~\cite{Finjord2007,BergmannPRE1995},
it is widely misused in the literature for other types of algorithms.
The results from NMR relaxation simulations are today therefore usually
fitted to experimental data with help of a superfluous free parameter.

Condition (3b) about the geometry is generally difficult, since we
do not know the true geometry of the media from which our digital
images have been extracted. We will use a simple linear interpolation
of the geometry, which is then merged into the boundary conditions
for surface relaxation. The general concept was introduced in~\cite{Ogren2014}
and was there described in detail for two spatial dimensions under
the name \emph{linear local boundary conditions} (LLBC).

In the present article we have combined the results of Eq.~(\ref{eq:p_S-rho-relation})
and the surface interpolation described in Sect.~\ref{sub:Local-Robin-boundary-conditions}
to treat the points (3a) and (3b) by constructing linear local Robin
boundary conditions to be used in Cartesian random walks in three-dimensions
and benchmarked their usefulness for modeling NMR time series in digital
media. Although interpolating digital media have been extensively
discussed for many years it does not appear to have been spread to
the NMR practitioners.

Regarding condition (1) above we have seen that the accuracy of the
obtained solution at any given time is dependent on the number of
walkers at $t=0$. The computational time of our software is roughly
linear in the number of trajectories, but the statistical standard
error in the total magnetisation is expected to scale as the square-root
of the number of trajectories.

All stochastically sampled data presented for three-dimensional samples
in this article is averaged over $10^{6}$ initial trajectories. For
the corresponding level of statistical accuracy the standard errors
are comparable to the linewidth of the curves presented in the figures
of the full time intervals (not for zoom-in figures, see e.g. Fig.~\ref{fig:Total-magnetization-of-the-ball}).

With the given hardware (an octa core workstation with 96 Gb RAM)
and software (Python code with the demanding subroutines in C) at
hand for this study, a simulation with $10^{6}$ trajectories walking
on $\sim10^{9}$ voxels the total computational time per sample was
in the order of $10$ hours. Hence, for example $\sim10^{3}$ trajectories
only take in the order of $10$ s to process, which is fast enough
for in-field analysis of geophysical samples. Consequently, the level
of statistical accuracy of the relaxation data generated from $10^{3}$
trajectories may be investigated in an ongoing applied project.

Regarding point (2) we used a computational lattice that agrees with
the resolution in the CT-images and with the cubic test domain we
got convergence to the correct solution (from Eq.~(\ref{eq:analytic_magnetization_for_cube}))
in the limit of $\Delta r\rightarrow0$.

\section{Benchmarking of the numerical method\label{sec:Tests-of-the-new-numerical-method}}

We start this Sect. with a qualitative discussion of the quantitative
results to be presented.

As shown in~\cite{Ogren2014}, the initial slope of the total magnetisation
in three-dimensions, both for the pore-domain $\Omega$ beeing a ball
of radius $R_{0}$ and a cube with sidelength $2R_{0}$, is given
by

\begin{equation}
\frac{1}{\mathcal{M}\left(0\right)}\left.\frac{d\mathcal{M}\left(t\right)}{dt}\right|_{t=0}=-\frac{3\rho}{R_{0}}.\label{eq:Asymptote_For_D_cube}
\end{equation}
Results of the type in Eq.~(\ref{eq:Asymptote_For_D_cube})
for the specific geometries presented here are generally valid for
any connected pore with a uniform initial condition~\cite{Ogren2014}
with the general result being

\begin{equation}
\frac{1}{\mathcal{M}\left(0\right)}\left.\frac{d\mathcal{M}\left(t\right)}{dt}\right|_{t=0}=-\rho\frac{S}{V}.\label{eq:Asymptote_rho_S_div_V}
\end{equation}
Here, $S$ represents the pore surface and $V$ the pore volume. Hence,
for the ball we have $S=4\pi R_{0}^{2}$ and $V=4\pi R_{0}^{3}/3$,
and for the cube $S=6\left(2R_{0}\right)^{2}$ and $V=\left(2R_{0}\right)^{3}$.
This explains Eq.~(\ref{eq:Asymptote_For_D_cube}) in those
cases. Now we can note that for fast diffusion (in relation to the
connectedness of the pore) the density is kept approximatelly uniform
and the result~(\ref{eq:Asymptote_rho_S_div_V}) holds for any time,
i.e., 
\begin{equation}
\mathcal{M}\left(t\right)\simeq\mathcal{M}\left(0\right)\mathrm{e}^{-\rho\frac{S}{V}t}.\label{eq:Approximation_for_constant_M}
\end{equation}
Many estimates based on the pore distribution function of a porous
media are based on the approximation~(\ref{eq:Approximation_for_constant_M})~\cite{MendelsonPRB1990}.
This is also in agreement with the interpretation of a lowest dominating
mode 
\begin{equation}
\mathcal{M}\left(t\right)\simeq\mathcal{M}\left(0\right)\mathrm{e}^{-\lambda_{1}t},\:\lambda_{1}\simeq\rho S/V,\label{eq:Approximation_lowest_dominating_mode}
\end{equation}
as discussed in~\cite{BrownsteinPRA1979}.

Generally a larger value of the surface relaxation parameter $\rho$
will cause the magnetic moment $M\left(\mathbf{x},t\right)$ to decay
more rapidly close to surfaces and more modes are needed to describe
$M\left(\mathbf{x},t\right)$. Hence the approximations~(\ref{eq:Approximation_for_constant_M})
and~(\ref{eq:Approximation_lowest_dominating_mode}) gradually fails.
There is also a qualitative dependence on the geometry of the pore-domain
since there is more surface area close to corners. Hence sharp corners
makes the magnetic moment more multiexponential.

With the true (lowest) eigenvalues, $\lambda_{j}$, for the ball and
cube with physical parameters such as $\rho S/V=3$ and $\rho=R_{0}=D_{0}=1$
reported in Table~\ref{Table_for_ball_and_cube}, we can compare
the approximation in~(\ref{eq:Approximation_lowest_dominating_mode})

\begin{equation}
\lambda_{1}^{ball}=2.4674\approx3,\:\lambda_{1}^{cube}=3\cdot0.74017=2.2205\approx3.
\end{equation}

\subsection{Comparison with FEM calculations in two-dimensional artificial porous
media\label{sub:Comparison-with-FEM_in_2D}}

\begin{figure}
\hspace{10mm}\textbf{(a)}\hspace{20mm}\textbf{(b)}\hspace{21mm}\textbf{(c)}

\resizebox{0.49\textwidth}{!}{%
  \includegraphics{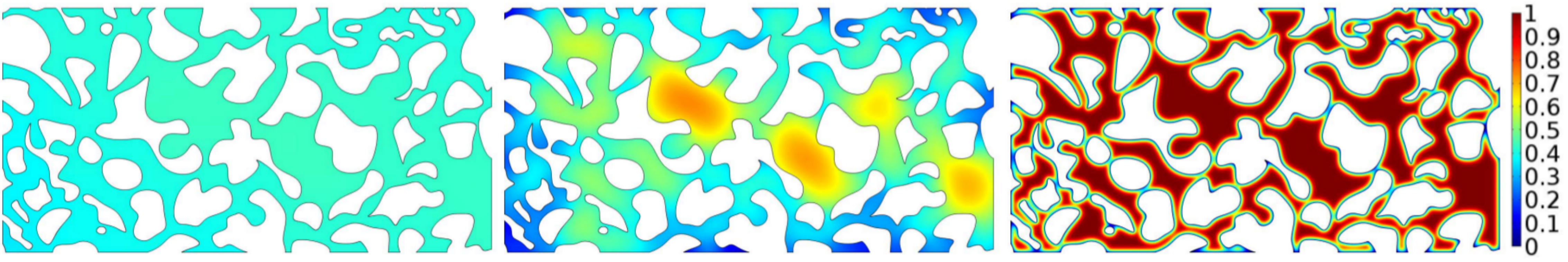}
}
\caption{Spatial patterns of the calculated relative NMR data, $M\left(\mathbf{x},t\right)/M\left(\mathbf{x},0\right)$,
in different parameter regimes at the time $t=1$ s obtained with
an ``extremely fine'' grid on the FEM software Comsol v. 4.2. (a)
Inter-pore \emph{fast} diffusion: the diffusion constant is here $D_{0}=10^{-7}$
m$^{2}$/s; (b) Intra-pore diffusion: $D_{0}=10^{-9}$ m$^{2}$/s;
(c) \emph{Slow} diffusion regime: $D_{0}=10^{-11}$ m$^{2}$/s. We
have kept the surface relaxation parameter fixed $\rho\simeq10^{-5}$
m/s in all cases, such that the ratio $T_{D_{0}}/T_{\rho}$ (see Eqs.~(\ref{eq:Diffusion-time})
and~(\ref{eq:Surface-relaxation-time})) is increasing by a factor
$100$ from (a) to (b), and from (b) to (c). \label{fig:Spatial-patterns-of-2D-artificial-porous-media}}
\end{figure}

For two-dimensional domains of limited size it is still doable to
use an extremely fine gridded finite element simulations (FEM) which
in effect interpolates the original digital domain. The complexity
of these calculations were relatively low and could be carried out
on a standard PC in the order of minutes. We therefore compare the
random walk method with and without the LLBC correction presented,
with FEM simulations for a two-dimensional artificial porous media
that is depicted in Figure~\ref{fig:Spatial-patterns-of-2D-artificial-porous-media}.
Examples of spatial patterns for the magnetic moment $M\left(\mathbf{x},t\right)$
for a certain time are presented in Figure~\ref{fig:Spatial-patterns-of-2D-artificial-porous-media}.
Subfigure (a) shows an example of fast diffusion where diffusion between
different pores happens \emph{faster} than surface relaxation, so
called inter-pore diffusion; subfigure (b) shows an intermediate example
where the diffusion is efficient only within the smaller pores, while
large (colour code yellow) pores still have a strong amplitude left;
finally in (c), the diffusion is so slow that in effect only the (one-dimensional)
surface have been affected by the relaxation at $t=1$. 

\begin{figure}

\resizebox{0.23\textwidth}{!}{
\includegraphics{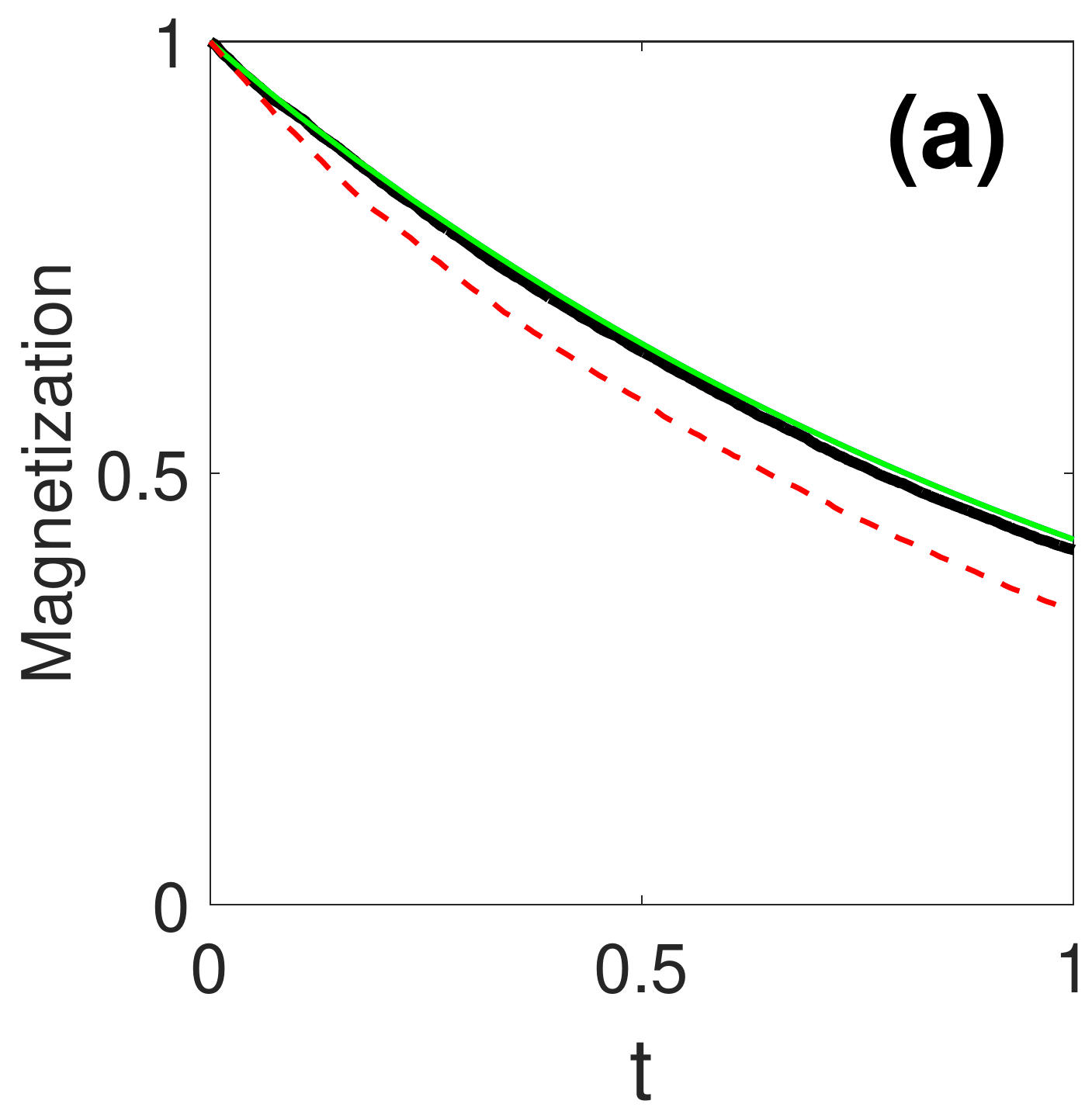}
}
\resizebox{0.23\textwidth}{!}{
\includegraphics{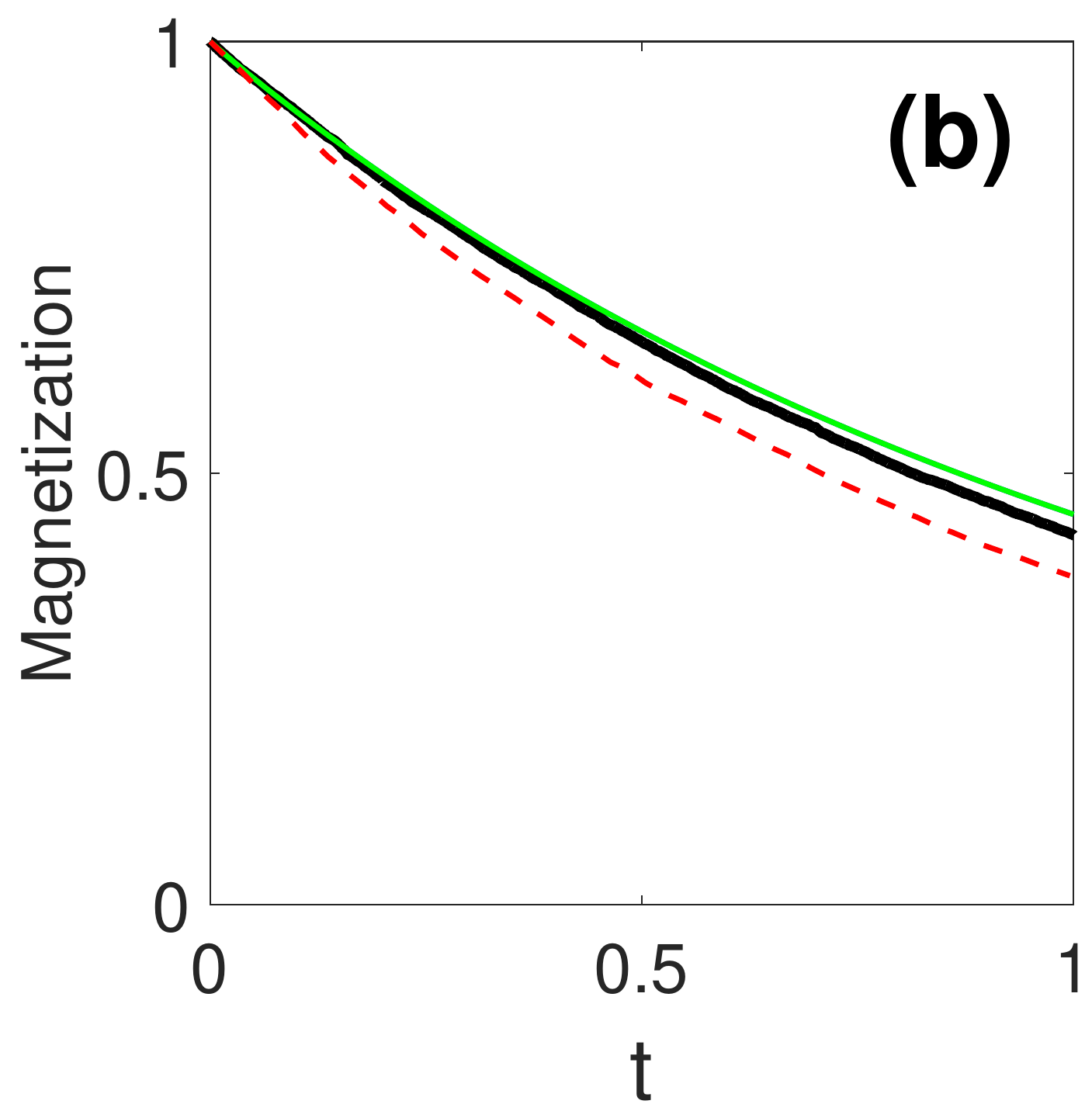}
}

\caption{Total magnetisation as a function of time. The upper (green) thin
curves are from FEM calculations with an ``extremely fine'' grid
in the FEM software Comsol v. 4.2. The lowest (red) dashed curves
are calculated with the original Cartesian random walk method for
two-dimensional domains, while the mid (black) thick curve is calculated
using the improved (LLBC) random walk method. The two-dimensional
media had $548\times274\sim10^{5}$ pixels with a resolution of $\Delta r=\Delta x=\Delta y=1.17\cdot10^{-6}$
m, hence the physical size of the media was in the order of $\sim0.5$
mm. A realistic parameter value for the diffusion constant in brine
is $D_{0}\simeq2\cdot10^{-9}$ m$^{2}$/s~\cite{Kenyon1992}, and
we show here two 2D examples with: Fig. (a) $D_{0}=10^{-9}$ m$^{2}$/s;
and Fig. (b) $D_{0}=10^{-8}$ m$^{2}$/s. The surface relaxation
parameter was kept fixed $\rho\simeq10^{-5}$ m/s~\cite{Vincent2011}
and no volume relaxation ($T_{V}\rightarrow\infty$) was present.
The number of initial trajectories is $10^{4}$ for this two-dimensional
geometry, which is enough for the standard deviation to be in the
order of the thickness of the linewidths in the plots. \label{fig:Timeseries-for-the-total-magnetization-of-2D-artificial-porous-media} }
\end{figure}

We can obtain the total magnetisation as a function of time, see Figure~\ref{fig:Timeseries-for-the-total-magnetization-of-2D-artificial-porous-media},
via Eq.~(\ref{eq:Def_of_total_magnetization}), equivalent to
integrating spatial patterns as in Figure~\ref{fig:Spatial-patterns-of-2D-artificial-porous-media}
for each time. It is clearly seen that the local boundary conditions
are superior in approximation the FEM results, that are here considered
as reference curves. 

\subsection{Benchmarking of the method for a ball and a cube\label{sub:Benchmarking-of-the-method-for-a-ball-and-a-cube}}

\begin{figure}

\resizebox{0.22\textwidth}{!}{
\includegraphics{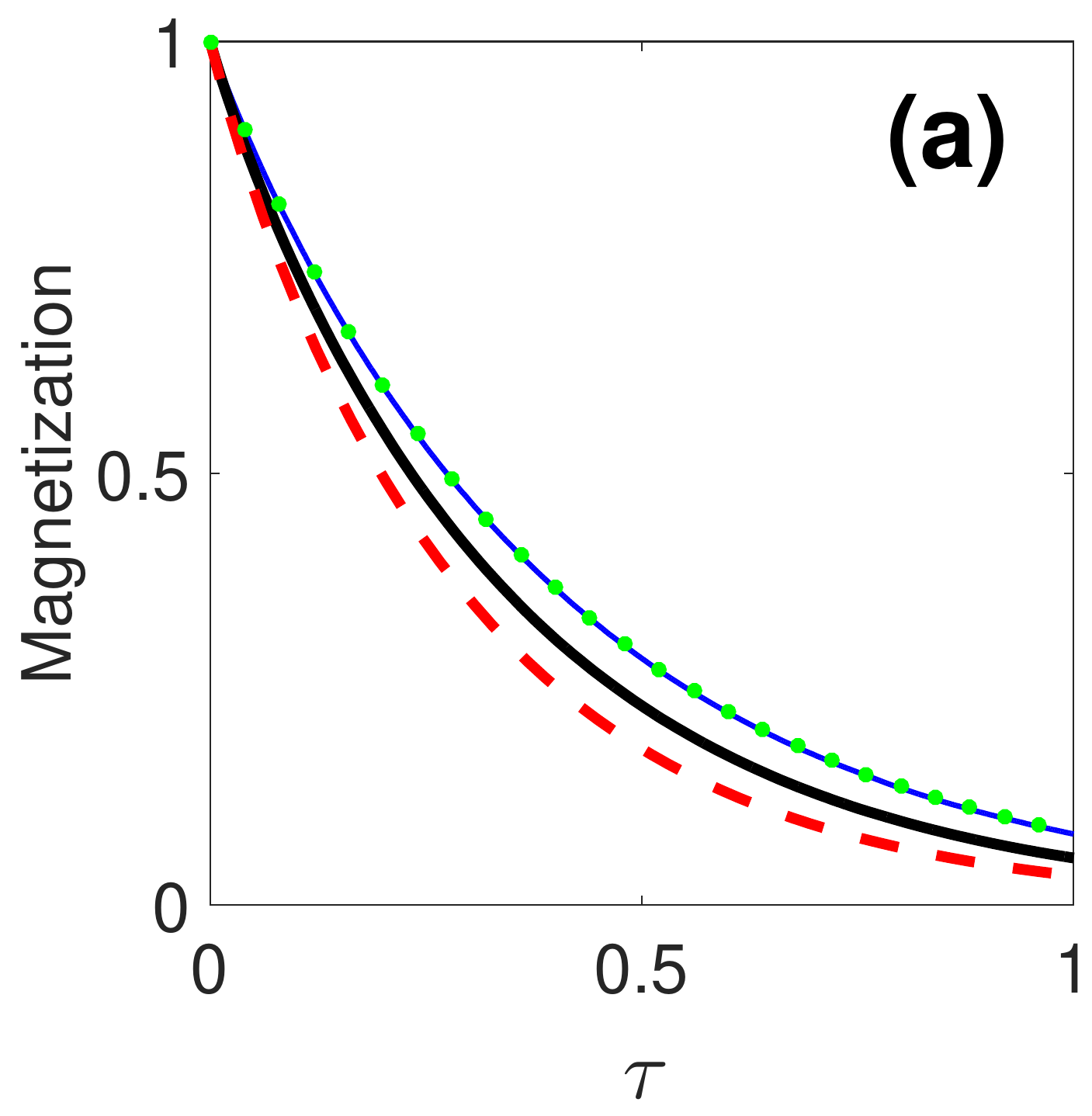}
}
\resizebox{0.24\textwidth}{!}{
\includegraphics{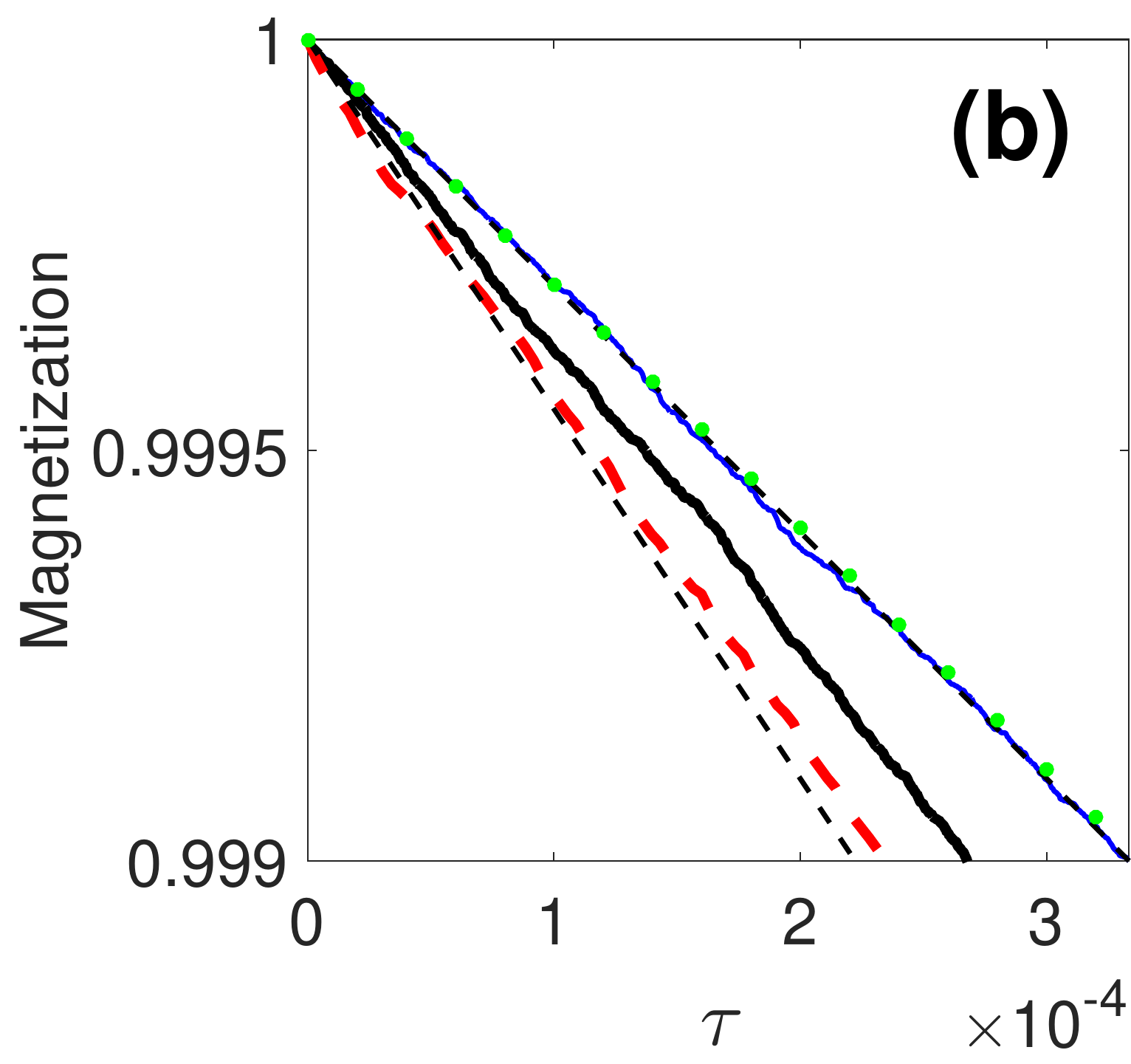}
}

\caption{Total magnetisation of the ball. In (a) the full time interval is
shown in the dimensionless time variable $\tau=D_{0}t/R_{0}^{2}$.
The (red) thick dashed curves show the result from the Cartesian random
walk with the original boundary condition. The (black) thick solid
curves are from the random walk with the LLBC improvement incorporated.
The analytic solution of Eq.~(\ref{eq:analytic_magnetization_for_ball}),
with $27$ terms included in the sum such that $1-\sum_{j}b_{j}\simeq10^{-6}$,
is shown with (green) dots. Solid (blue) curves show the result obtained
with a radial random walk~\cite{Ogren2014}. In (b) we show in addition
two analytic short time asymptotes as (black) thin lines. The upper
right line has the slope $-3\rho/R_{0}$ (according to Eq.~(\ref{eq:Asymptote_For_D_cube})),
while the lower left line has the slope $-9\rho/\left(2R_{0}\right)=-\rho\cdot6\cdot\pi R_{0}^{2}/\left(4\pi R_{0}^{3}/3\right)$.\label{fig:Total-magnetization-of-the-ball}}
\end{figure}

In three dimensions we have first tested our implemention of the LLBC
by benchmarking against analytic solutions for simple geometries,
as was done for two dimensions in~\cite{Ogren2014}. As the first
analytic reference, we present a series expansion of the solution
for the total magnetisation (with $T_{V}\rightarrow\infty$, see Eq.~(\ref{eq:T_V_factorisation}))
of a ball with uniform initial conditions, i.e., $M\left(\mathbf{x},0\right)=M_{0}$~\cite{BrownsteinPRA1979}
giving
$$
\mathcal{M}\left(t\right)=12\mathcal{M}\left(0\right)
$$
\begin{equation}
\times \sum_{j=1}^{\infty}\frac{\left[\sin\left(\sqrt{\lambda_{j}}\right)-\sqrt{\lambda_{j}}\cos\left(\sqrt{\lambda_{j}}\right)\right]^{2}}{\lambda_{j}^{3/2}\left[2\sqrt{\lambda_{j}}-\sin\left(2\sqrt{\lambda_{j}}\right)\right]}\mathrm{e}^{-\frac{D_{0}}{R_{0}^{2}}\lambda_{j}t},\label{eq:analytic_magnetization_for_ball}
\end{equation}
where the eigenvalues $\lambda_{j}$ are solutions of the equation
$$1-\sqrt{\lambda_{j}}\cot\left(\sqrt{\lambda_{j}}\right)=R_{0}\rho/D_{0}.$$
In Figure~\ref{fig:Total-magnetization-of-the-ball} we compare numerical
solutions with and without LLBC to the analytic solution of Eq.~(\ref{eq:analytic_magnetization_for_ball}).
We see that LLBC improves the result for the ball, while a systematic
deviation is still present.

To further illustrate the dramatic effect of the digitized surface
of the ball (sphere) we have also plotted numerical results obtained
from a radial random walk, which is the same data as the ``$D=3$''-curve
in Figure 1 (a) of~\cite{Ogren2014}, that converge to the analytic
solution given a small enough discretization $\Delta r$ of the radius
of the ball.

\begin{figure}
\resizebox{0.22\textwidth}{!}{
\includegraphics{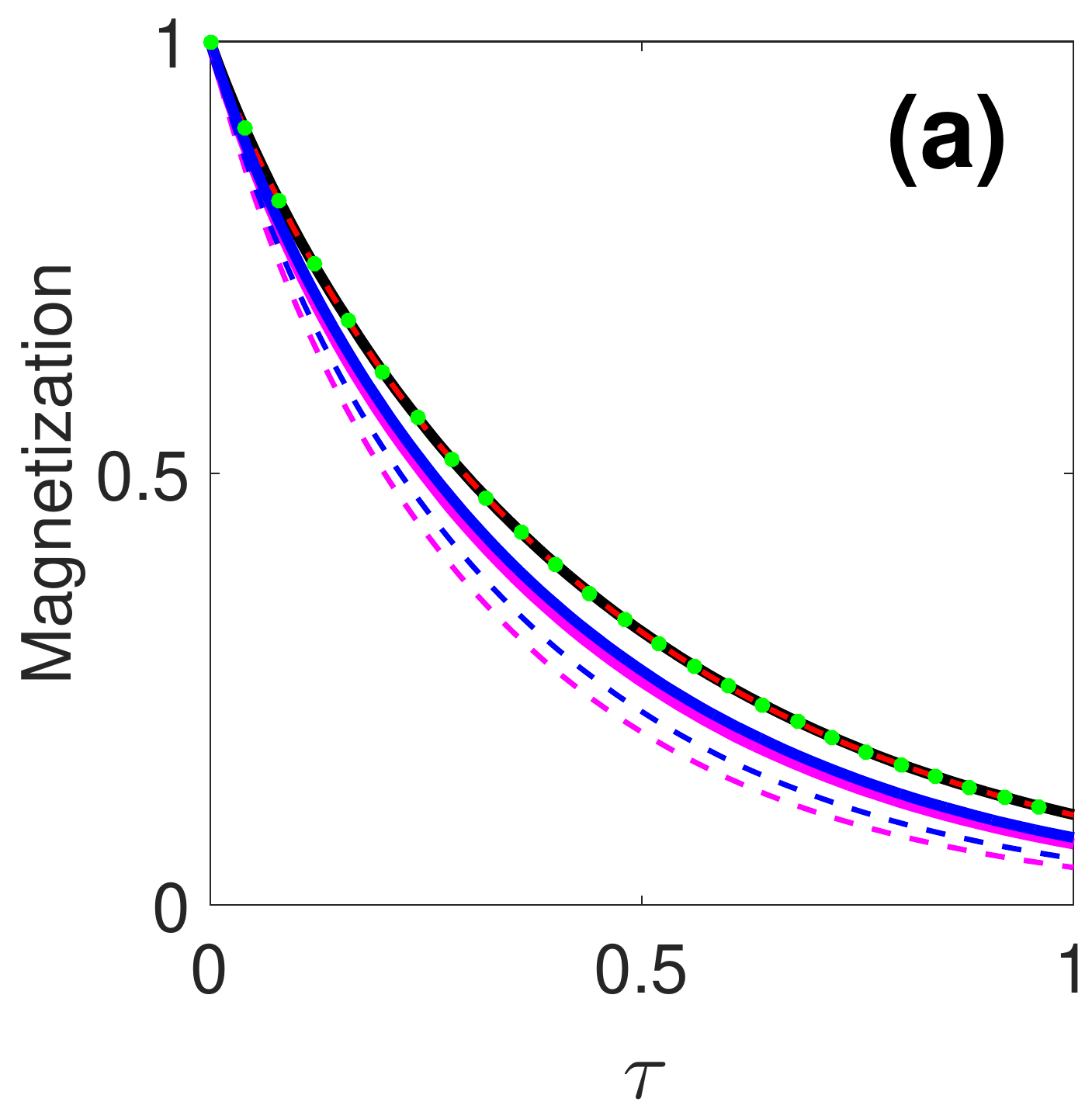}
}
\resizebox{0.24\textwidth}{!}{
\includegraphics{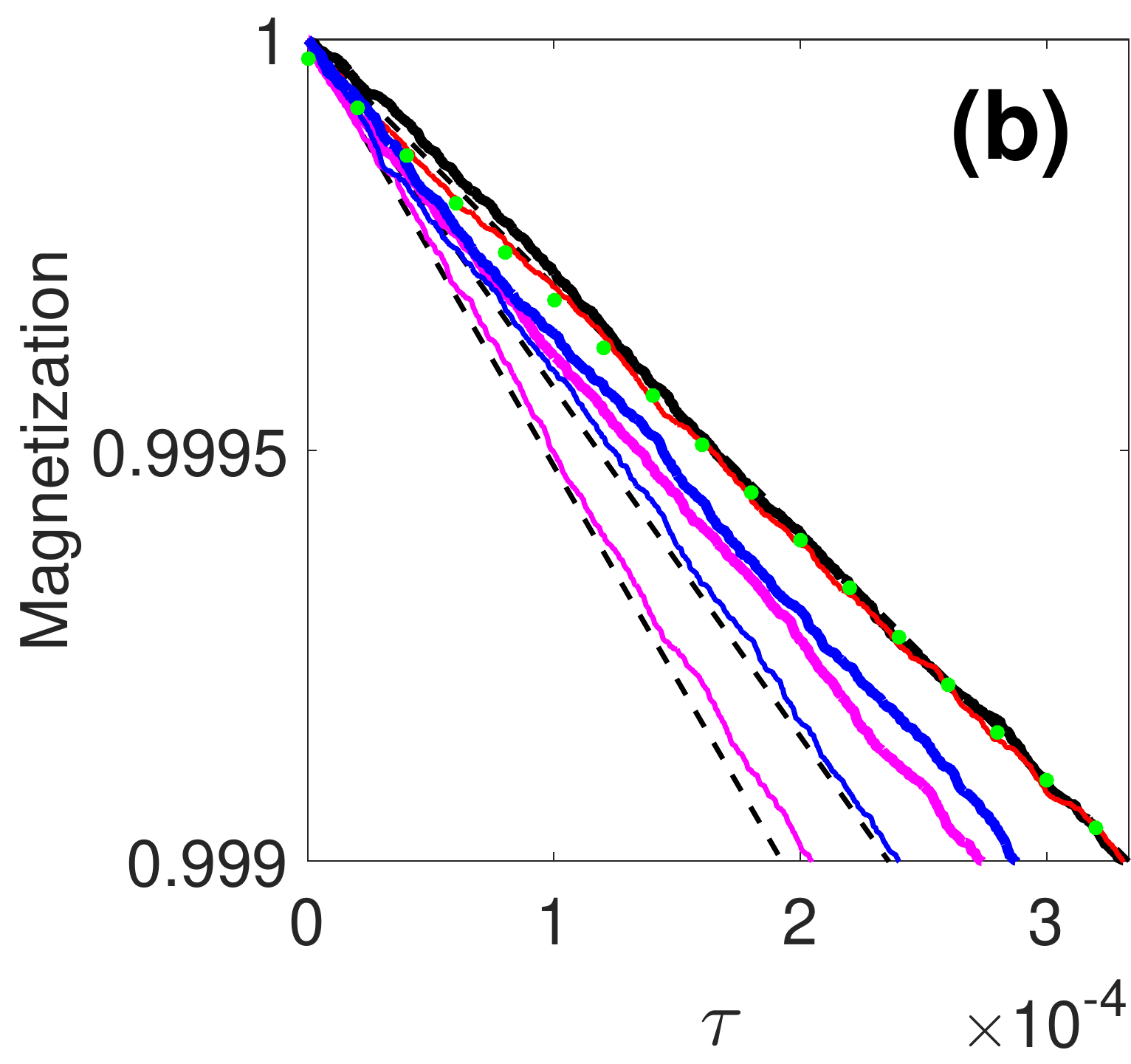}
}
\caption{Total magnetisation of the cube in different orientations. (a) shows
the full time interval in the dimensionless time variable $\tau=D_{0}t/R_{0}^{2}$.
The (red) thick dashed curves shows the result from the Cartesian
random walk with the original boundary condition. The (black) thick
solid curves are from the random walk with the LLBC improvement incorporated.
The blue pair of curves, dashed line for original boundary condition,
thick solid line for the LLBC incorporated, shows the result for a
cube rotated $\left(\pi/8,\:\pi/8\right)$. The magenta pair of curves
shows the result for a cube rotated $\left(\pi/4,\:\pi/4\right)$.
The analytic solution of Eq.~(\ref{eq:analytic_magnetization_for_cube}),
with $29$ terms included in the sum such that $1-2\sum_{j}c_{j}\simeq10^{-6}$,
is shown with (green) dots. (b) shows in addition three analytic short
time asymptotes as (black) thin dashed lines. The upper right dashed
line (covered by other symbols) has the slope $-3\rho/R_{0}$ (according
to Eq.~(\ref{eq:Asymptote_For_D_cube})), while the mid dashed line
has the slope $-3\sqrt{2}\rho/R_{0}$ and the lower left dashed line
has the slope $-3\sqrt{3}\rho/R_{0}$.}
\label{fig:Total-magnetization-of-the-cube} 
\end{figure}

The second of the two simple three-dimensional geometries we consider
is the cube. While the results of the $D$-dimensional case were derived
for dimensionless variables in~\cite{Ogren2014} for different initial
conditions with help of Sturm-Liouville theory, we here give explicitly
the total magnetisation for the cube with uniform initial conditions
$$
\mathcal{M}\left(t\right)=8\mathcal{M}\left(0\right)
$$
\begin{equation}
\times \left[\sum_{j=1}^{\infty}\frac{\sin^{2}\left(\sqrt{\lambda_{j}}\right)}{\sqrt{\lambda_{j}}\sin\left(\sqrt{\lambda_{j}}\right)\cos\left(\sqrt{\lambda_{j}}\right)+\lambda_{j}}\mathrm{e}^{-\frac{D_{0}}{R_{0}^{2}}\lambda_{j}t}\right]^{3}.\label{eq:analytic_magnetization_for_cube}
\end{equation}
Now the eigenvalues $\lambda_{j}$ are solutions of the equation 
$$\sqrt{\lambda_{j}}\tan\left(\sqrt{\lambda_{j}}\right)=R_{0}\rho/D_{0}.$$
In Figure~\ref{fig:Total-magnetization-of-the-cube} we compare numerical
solutions with and without LLBC to the analytic solution of Eq.~(\ref{eq:analytic_magnetization_for_cube}).
For the (non-rotated) cube the LLBC have no effect since the result
is already exact. When rotating the cube, with respect to the coordinate
axis, large deviations occurs (dashed curves) that are only partly
corrected with LLBC.

For the discussion to follow and the convenience of the readers in
checking their own codes for NMR relaxation we include the necessary
numbers in Table~\ref{Table_for_ball_and_cube} in order to use the
formulas of Eqs.~(\ref{eq:analytic_magnetization_for_ball})
and~(\ref{eq:analytic_magnetization_for_cube}) in practice. From
Eq.~(\ref{eq:analytic_magnetization_for_ball}) we define $b_{j}$
as the coefficients in the series for the ball ($b$), while for Eq.~(\ref{eq:analytic_magnetization_for_cube})
we define $c_{j}=\sin^{2}\left(\sqrt{\lambda_{j}}\right)/\left[\sqrt{\lambda_{j}}\sin\left(\sqrt{\lambda_{j}}\right)\cos\left(\sqrt{\lambda_{j}}\right)+\lambda_{j}\right]$
as the terms in the corresponding series $$\mathcal{M}\left(0\right)^{1/3}=\mathcal{M}\left(0\right)^{1/3}2\sum_{j=1}^{\infty}c_{j}=1$$
for the cube ($c$), see Table~\ref{Table_for_ball_and_cube}.
\begin{table}
\scriptsize 
\begin{tabular}{|c|c|c|c|c|}
\hline 
$j$  & $\lambda_{j}$ ball (\ref{eq:analytic_magnetization_for_ball})  & $b_{j}$ ball  & $\lambda_{j}$ cube (\ref{eq:analytic_magnetization_for_cube})  & $c_{j}$ cube\tabularnewline
\hline 
\hline 
$1$  & $2.4674$  & $0.98553$  & $0.74017$  & $0.49305$\tabularnewline
\hline 
$2$  & $22.207$  & $0.012167$  & $11.735$  & $6.2044\cdot10^{-3}$\tabularnewline
\hline 
$3$  & $61.685$  & $1.5769\cdot10^{-3}$  & $41.439$  & $5.5554\cdot10^{-4}$\tabularnewline
\hline 
$4$  & $120.90$  & $4.1047\cdot10^{-4}$  & $90.808$  & $1.1866\cdot10^{-4}$\tabularnewline
\hline 
$5$  & $199.86$  & $1.5021\cdot10^{-4}$  & $159.90$  & $3.8627\cdot10^{-5}$\tabularnewline
\hline 
$6$  & $298.56$  & $6.7313\cdot10^{-5}$  & $248.73$  & $1.6034\cdot10^{-5}$\tabularnewline
\hline 
$7$  & $416.99$  & $3.4506\cdot10^{-5}$  & $357.30$  & $7.7895\cdot10^{-6}$\tabularnewline
\hline 
$8$  & $555.17$  & $1.9467\cdot10^{-5}$  & $485.61$  & $4.2232\cdot10^{-6}$\tabularnewline
\hline 
$9$  & $713.08$  & $1.1800\cdot10^{-5}$  & $633.65$  & $2.4827\cdot10^{-6}$\tabularnewline
\hline 
$10$  & $890.73$  & $7.5624\cdot10^{-6}$  & $801.44$  & $1.5530\cdot10^{-6}$\tabularnewline
\hline 
\end{tabular}
\caption{\textit{For the 3D test cases presented in Figures~\ref{fig:Total-magnetization-of-the-ball}
and~\ref{fig:Total-magnetization-of-the-cube} we here give the first
$10$ eigenvalues and corresponding coefficients for the parameter
values $\rho=R_{0}=D_{0}=1$ (i.e. intermediate diffusion regime).
Note the qualitative difference in the values for the ball and cube
for a given $j$. This is due to the difference in the definition
of the series, see Eqs.~(\ref{eq:analytic_magnetization_for_ball})
and~(\ref{eq:analytic_magnetization_for_cube}) respectively.} }
\label{Table_for_ball_and_cube} 
\end{table}

\section{NMR relaxation in large complex domains\label{sec:Application-to-NMR-relaxation-in-true-porous-media}}

After having benchmarked LLBC with an artificial 2D porous media in
Sect.~\ref{sub:Comparison-with-FEM_in_2D}, and with the analytic
solutions of the ball and the cube in 3D in Sect.~\ref{sub:Benchmarking-of-the-method-for-a-ball-and-a-cube},
we now apply LLBC to large complex digital domains from CT-images
of chalk.

\subsection{Complex geometries, digital images of chalk\label{sub:Complex-geometries,-digital-images-of-chalk}}

\begin{figure}
\hspace{16mm}\textbf{\large{(a)}}{\Large{\hspace{36mm}}}\textbf{\large{(b)}}{\Large{\vspace{1.5mm}
 }}{\Large \par}
\resizebox{0.49\textwidth}{!}{
\includegraphics{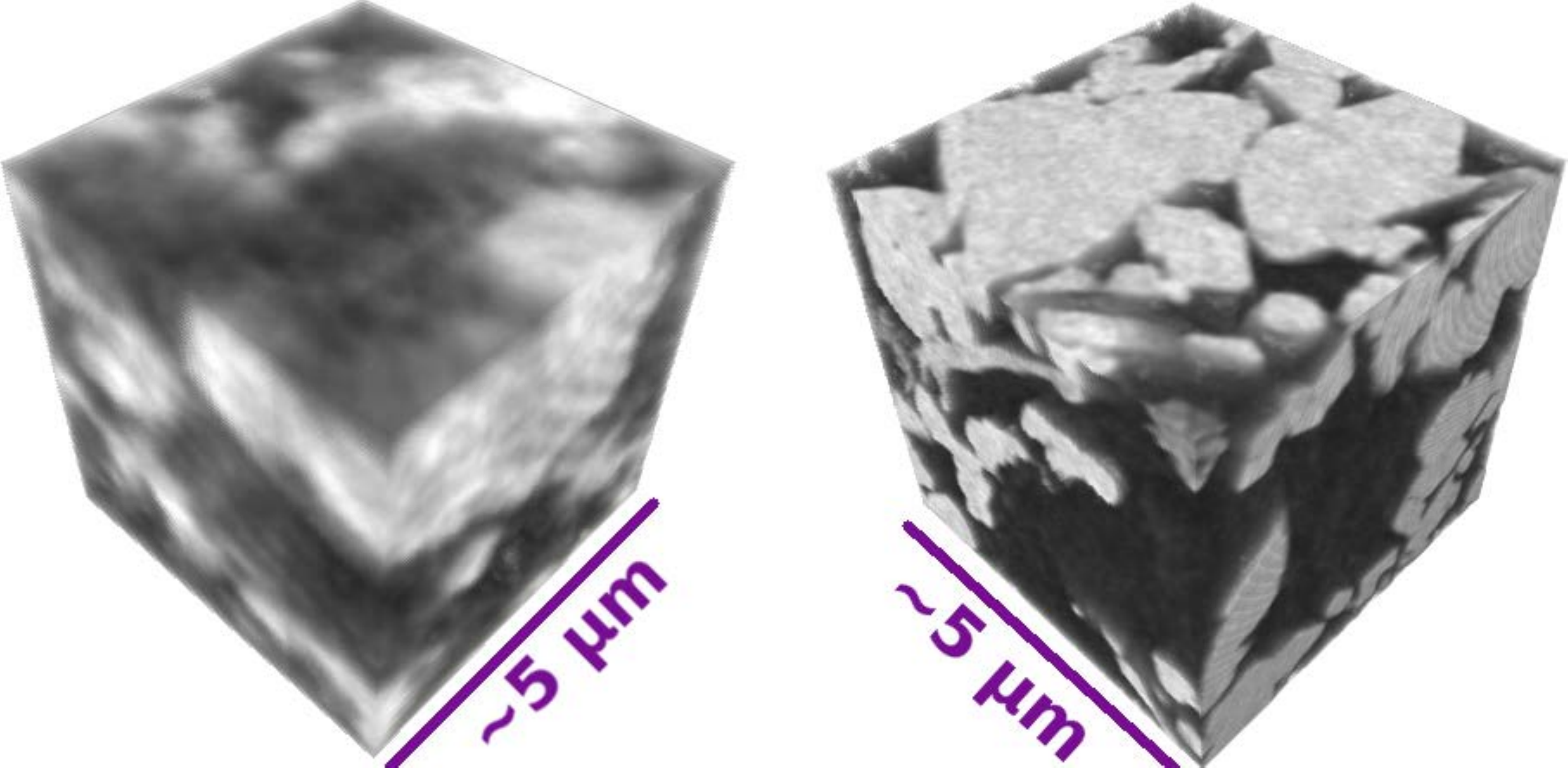}
}
\caption{Examples of zooming in on regions in the CT-images for the two 3D
chalk samples. To the left limestone (a) and to the right Aalborg
chalk (b). Black ($Z=1$) voxels constitutes the pore domain (assumed
to be filled with brine). White ($Z=0$) voxels represents regions
of homogenous chalk. Aalborg chalk (b) contains a more complex pore
morphology and was imaged with ptychographic X-ray nanotomography,
which explains the sharper image compared to (a). Since the side length
of the full samples are in the order of $N\Delta r\simeq25$ $\mu$m
(Table~\ref{Table_for_real_samples}), the subvolumes shown here
represents only about $1\%$ of the chalk sample domains in our calculations.
\label{fig:3D_CT-images}}
\end{figure}

A limestone sample was taken from an outcrop at Rødvig (Stevns Klint)
in Denmark. A subsample ($\sim500$ $\mu$m in diameter) of this
sample was imaged using the holotomography setup at the former ID22~\cite{koch1998}
at the European Synchrotron Radiation Facility in Grenoble, France
at four resolutions: $320$, $100$, $50$ and $25$ nm voxel size~\cite{mueter2014}.
For the purpose of this work, we have chosen the data set at $25$
nm voxel size to ensure the best possible result. As an example of
a very complex pore geometry, we collected samples of chalk from a
quarry in Aalborg, Denmark. 
Using a focused gallium ion beam in a
scanning electron microscope, we produced cylindrical samples of $\sim20$
$\mu$m in diameter fit for imaging using the recently developed
ptychography method. Imaging itself was performed at the Swiss Light
Source in Villigen, Switzerland at the cSAXS beamline~\cite{holler2014}
resulting in a voxel size of $21.47$ nm. Basic properties for the
two digital domains from the two chalk samples, in part illustrated
in Figure~\ref{fig:3D_CT-images}, are presented in Table~\ref{Table_for_real_samples}.
\begin{table}
\tiny 
\begin{tabular}{|c|c|c|c|c|}
\hline 
Property  & Aalborg  & Aal., LLBC  & Limestone  & Lim., LLBC\tabularnewline
\hline 
\hline 
$N$  & $696$  & $696$  & $1279$  & $1279$\tabularnewline
\hline 
$\Delta r$ {[}nm{]}  & $21.47$  & $21.47$  & $25$  & $25$\tabularnewline
\hline 
$V$ {[}$\left(\Delta r\right)^{3}${]}  & $156789057$  & $156789057$  & $1043737196$  & $1043737196$\tabularnewline
\hline 
$S$ {[}$\left(\Delta r\right)^{2}${]}  & $27291763$  & $20738034$  & $61403293$  & $48152708$\tabularnewline
\hline 
$\phi=\frac{V}{N^{3}}$  & $0.465$  & $0.465$  & $0.499$  & $0.499$\tabularnewline
\hline 
$SSA=\frac{S}{N^{3}}$ {[}$\left(\Delta r\right)^{-1}${]}  & $0.0809$  & $0.0615$  & $0.0293$  & $0.0230$\tabularnewline
\hline 
$SSA$ {[}m$^{2}$/g{]}  & $1.4$  & $1.1$  & $0.43$  & $0.34$\tabularnewline
\hline 
\end{tabular}
\caption{\textit{Data for the two real samples described in Sect.~\ref{sub:Complex-geometries,-digital-images-of-chalk}.
Here $V$ is the pore volume, $S$ the pore surface, and $\phi$ the
porosity of the digital domains. The specific surface area (SSA) in
the last row has been estimated using a density of $2.7$~g/cm$^{3}$
for calcite, which is the dominant compound of both samples~\cite{soerensen2012}.} }
\label{Table_for_real_samples} 
\end{table}

The corresponding NMR relaxation for Aalborg chalk and limestone are
seen in Figure~\ref{fig:real_samples}. 

Our first sample, Aalborg chalk, was represented by a volume of $696^{3}$
voxels. The second sample, limestone, had $1279^{3}$ voxels. In order
to test the ability to later analyse even larger experimental samples,
we have treated artificial porous media with more than $\sim10^{10}$
voxels (i.e. $\sim10$ times the numbers of the real data in this
study) on a single PC. Due to the independence of individual trajectories,
it is in effect the RAM memory that limits the size of the sample
one can efficiently handle with a given hardware.

\subsection{Results of the total magnetisation in chalk\label{sub:Results-of-the-total-magnetization-in-chalk}}

\begin{figure}
\resizebox{0.22\textwidth}{!}{
\includegraphics{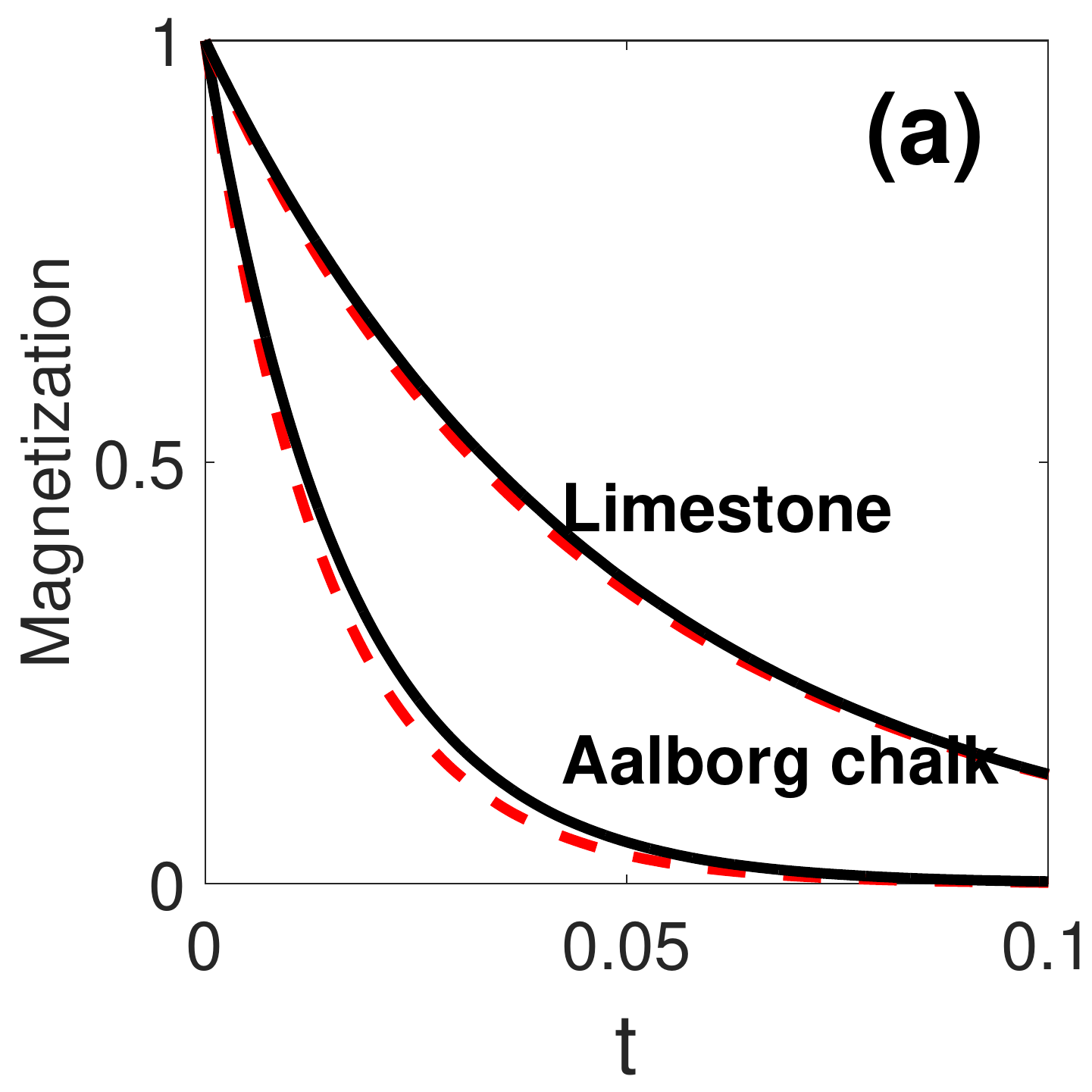}
}
\resizebox{0.24\textwidth}{!}{
\includegraphics{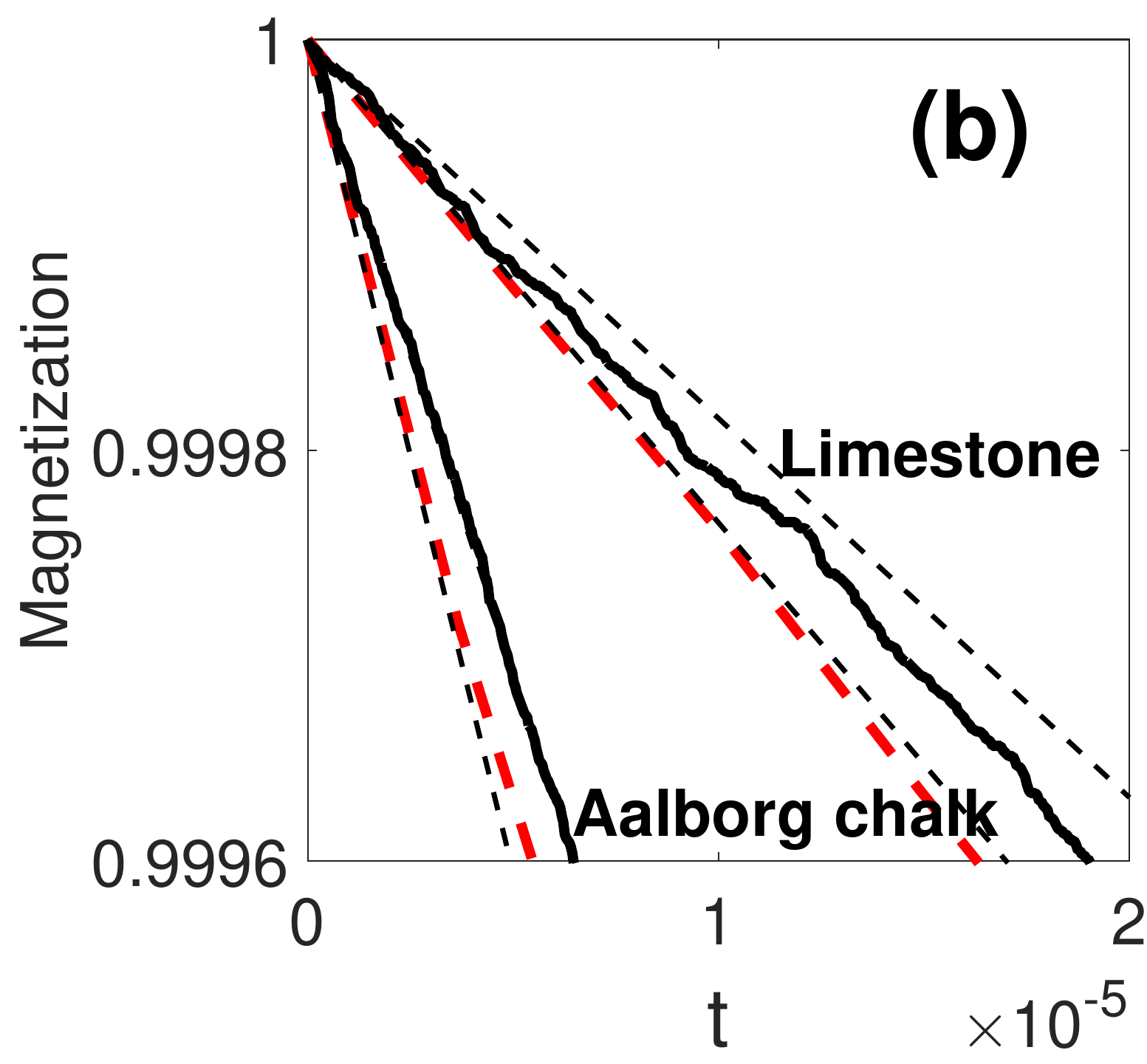}
}
\caption{Total magnetisation of the chalk samples, described in Sect.~\ref{sub:Complex-geometries,-digital-images-of-chalk},
for the relevant full time interval in seconds in (a). A zoom in for
a comparison with the analytic short time asymptotes in (b). The (red)
thick dashed curves show the result from the Cartesian random walk
with the original boundary condition. The (black) thick solid curves
are from the random walk with the LLBC improvement incorporated. In
(b) four analytic short time asymptotes are shown as (black) thin
dashed lines. The asymptotes all have the slope $-\rho S/V$ (according
to Eq.~(\ref{eq:Asymptote_rho_S_div_V})), while the ratio of the
pore surface $S$ and the pore volume $V$ varies with the sample
and the boundary condition in use, see Table~\ref{Table_for_real_samples}.
\label{fig:real_samples}}
\end{figure}

The results of applying the numerical methods of Sect.~\ref{sec:Method}
to the chalk samples are presented in Figure~\ref{fig:real_samples}.
We can observe qualitatively different results for the NMR relaxation
curves of the Aalborg chalk and the limestone samples. In fact their
half times, $T_{1/2}$, differ by more than a factor of two (see Fig.~\ref{fig:real_samples}
(a)). The half times are in the order of $T_{1/2}\sim10$ ms for the
Aalborg chalk and of $T_{1/2}\sim30$ ms for the limestone. According
to Table~\ref{Table_for_real_samples} the two samples have similar
porosity but quite different specific surface area. We interpret the
faster relaxation for the Aalborg chalk as mainly a consequence of
its richer surface structure and narrower pore throats. From Figure~\ref{fig:3D_CT-images},
we can estimate that most pores are of size much smaller than of $10$
$\mu$m. Hence, the discussion under Eqs.~(\ref{eq:Diffusion-time})
and~(\ref{eq:Surface-relaxation-time}) suggest that we are in the
regime of fast diffusion for both samples. Then we can expect the
approximation of Eq.~(\ref{eq:Approximation_for_constant_M})
to be relevant. Indeed we can qualitatively obtain the half times
as $T_{1/2}\sim\ln2\cdot V/\left(\rho S\right)\simeq8.5-11$ ms for
the Aalborg chalk and of $T_{1/2}\simeq29-38$ ms for the limestone
(the ranges are presented due to the two different columns for each
sample in Table~\ref{Table_for_real_samples}). However, by plotting
the curves corresponding to Eq.~(\ref{eq:Approximation_for_constant_M})
in the same graph as the numerically obtained relaxation curves (Fig.~\ref{fig:real_samples})
important quantitative differences shows up, so a more accurate description
is necessary for a quantitative analysis.

As mentioned earlier (e.g.) for brine we have in addition that the
volume relaxation time is $T_{V}\simeq3.1$ s and the true (experimental)
relaxation curves for the chalk samples are as those in Figure~\ref{fig:real_samples}
but multiplied with the common factor $\mathrm{e}^{-t/T_{V}}$ (see
Eq.~(\ref{eq:T_V_factorisation})). This lower the half times according
to $T_{1/2}\sim\ln2/\left(\rho S/V+1/T_{V}\right)$, which is only
about 1\% shorter than the half times stated above.

\subsection{Decomposition of the total magnetisation into Laplace components}

By a mathematical inversion process on the NMR relaxation data, it
is possible to obtain the familiar $T$-distribution curve, which
reflects the distribution of the pore surface-to-volume ratio of the
media~\cite{Vincent2011}. 

The starting point for presenting our method for a signal analysis
of the total magnetisation is the Laplace transform of the real decomposition
$F\left(\beta\right)$

\begin{equation}
\mathcal{M}\left(t\right)=\int_{0}^{\infty}F\left(\beta\right)\mathrm{e}^{-\beta t}d\beta.\label{eq:Laplace_transform}
\end{equation}

Finding $F$ given $\mathcal{M}$ is an ill-posed inverse problem
\cite{Berman2013}. Most commonly used methods retrieve a continuous
approximation for the distribution $F\left(\beta\right)$, a so called
$T$-distribution curve, using some sort of regularization to pick
an $F$ with small support. Alternatively, one may use the complex
frequency estimation technique ESPRIT~\cite{Roy1989} in order to
find few coefficients $d_{j}>0$ and corresponding inverse times $\beta_{j}=1/T_{j}>0$,
such that the distribution 
\begin{equation}
F\left(\beta\right)=\sum_{j}d_{j}\delta\left(\beta-\beta_{j}\right),
\end{equation}
via the Laplace transform~(\ref{eq:Laplace_transform}), gives an
approximation of the total magnetisation. In other words we here seek
representations of the form 
\begin{equation}
\mathcal{M}\left(t\right)\simeq\sum_{j=1}^{m}d_{j}\mathrm{e}^{-t/T_{j}}.\label{eq:few_terms_representation}
\end{equation}
Given that $\mathcal{M}$ is of the form~(\ref{eq:few_terms_representation}),
ESPRIT is guaranteed to find the precise parameters. However, in the
presence of noise, this may not be the case, and it can then be beneficial
to use more advanced techniques to pretreat the data in order to avoid,
e.g., complex frequency components. The algorithm presented in~\cite{Andersson_Eusipco_2016}
is tailormade for this purpose, which aims to minimize the $L^{2}$-error
\begin{equation}
\mathcal{N}=\sqrt{\int_{0}^{t_{2}}\left|\mathcal{M}\left(t\right)-\sum_{j=1}^{m}d_{j}\mathrm{e}^{-t/T_{j}}\right|^{2}dt},\label{eq:L2-norm}
\end{equation}
while at the same time enforcing the parameters $T_{j}$ to be real.
We here refer to this method as ``real exponents ESPRIT''.

We also present the results from a traditional method, using Tikhonov
Regularized Inversion (TRI)~\cite{ParkerJMR2005}. The regularization
parameter in TRI was chosen according to the discrepancy principle~\cite{PCH1998}
with an estimated error level of $10^{-3}$, obtained from a comparison
between a numerical and an analytic solution of the cube.
\begin{figure}
\resizebox{0.22\textwidth}{!}{
\includegraphics{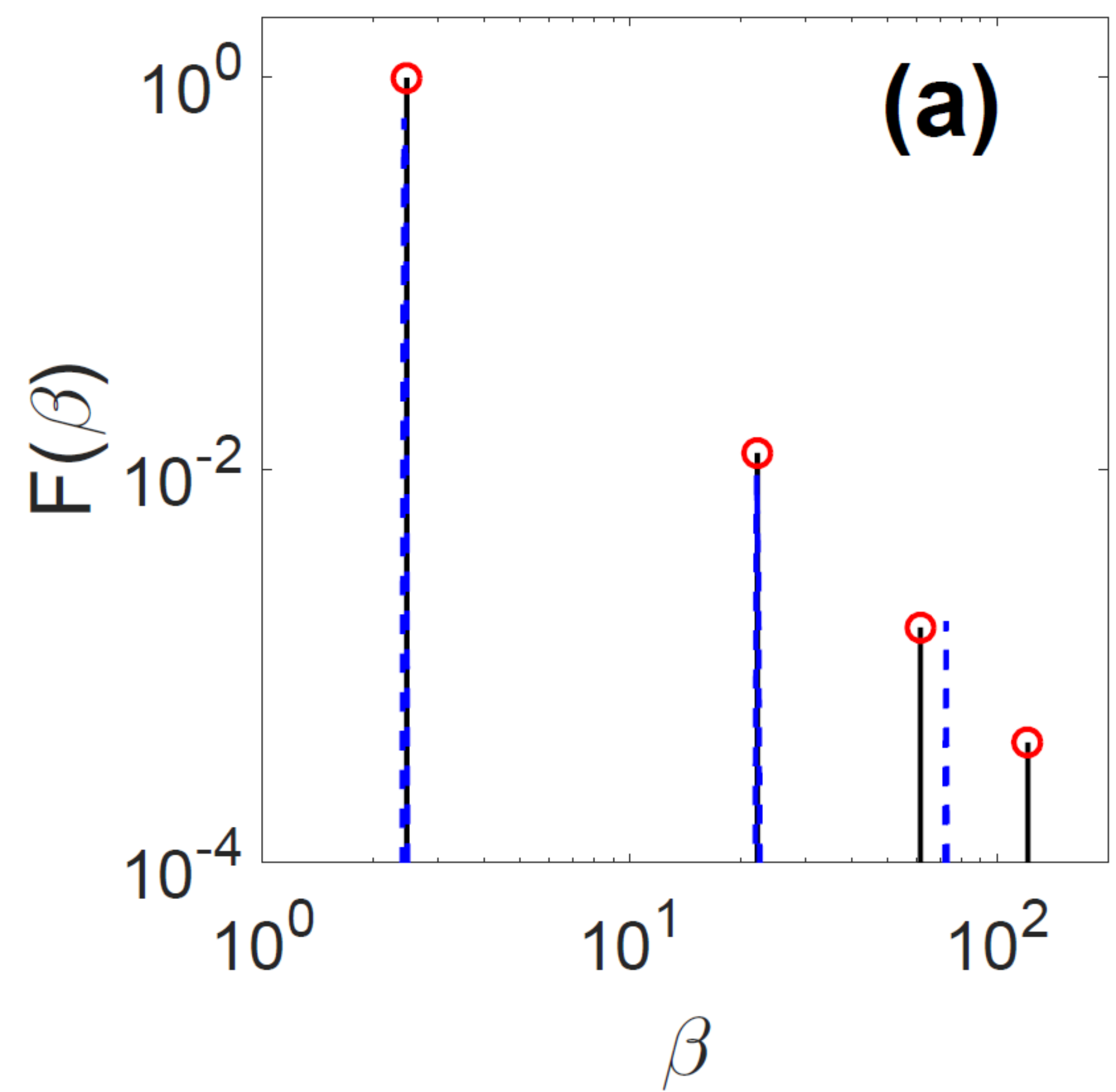}
}
\resizebox{0.25\textwidth}{!}{
\includegraphics{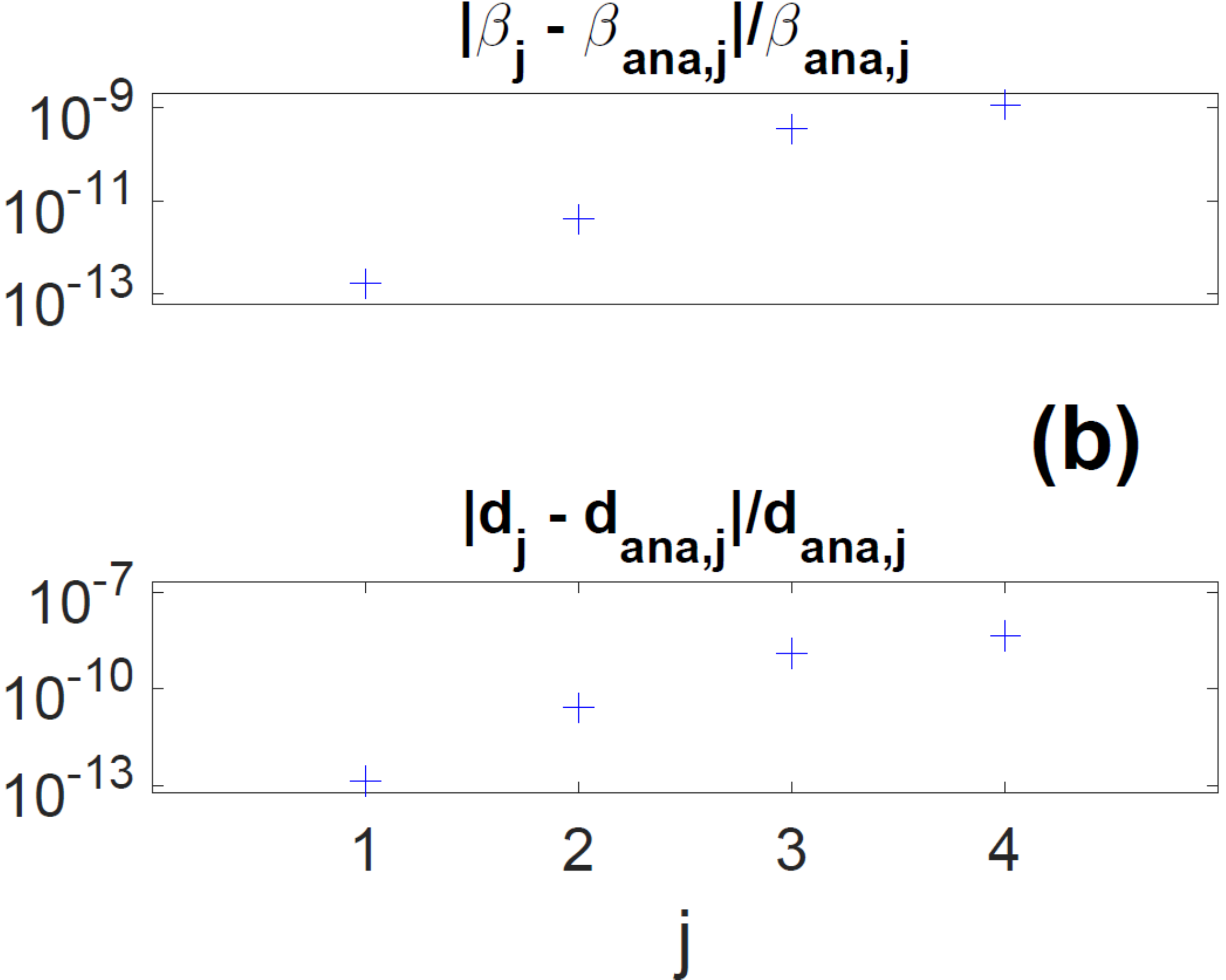}
}
\caption{Decomposition of the analytic total magnetisation of the ball, using
four terms in Eq.~(\ref{eq:analytic_magnetization_for_ball}).
(a) Loglog plot of numerically obtained coefficients and exponents,
vertical (black) lines shows the discrete result from ESPRIT, while
the continuous (blue) dashed-curve are from the Tikhonov regularized
inversion. The original four analytic datapoints, see Table~\ref{Table_for_ball_and_cube},
are shown as (red) open circles. (b) Relative deviation for the four
numerically obtained data points in (a) using ESPRIT. \label{fig:decomposition_of_analytic_ball}}
\end{figure}

\begin{figure}

\resizebox{0.23\textwidth}{!}{
\includegraphics{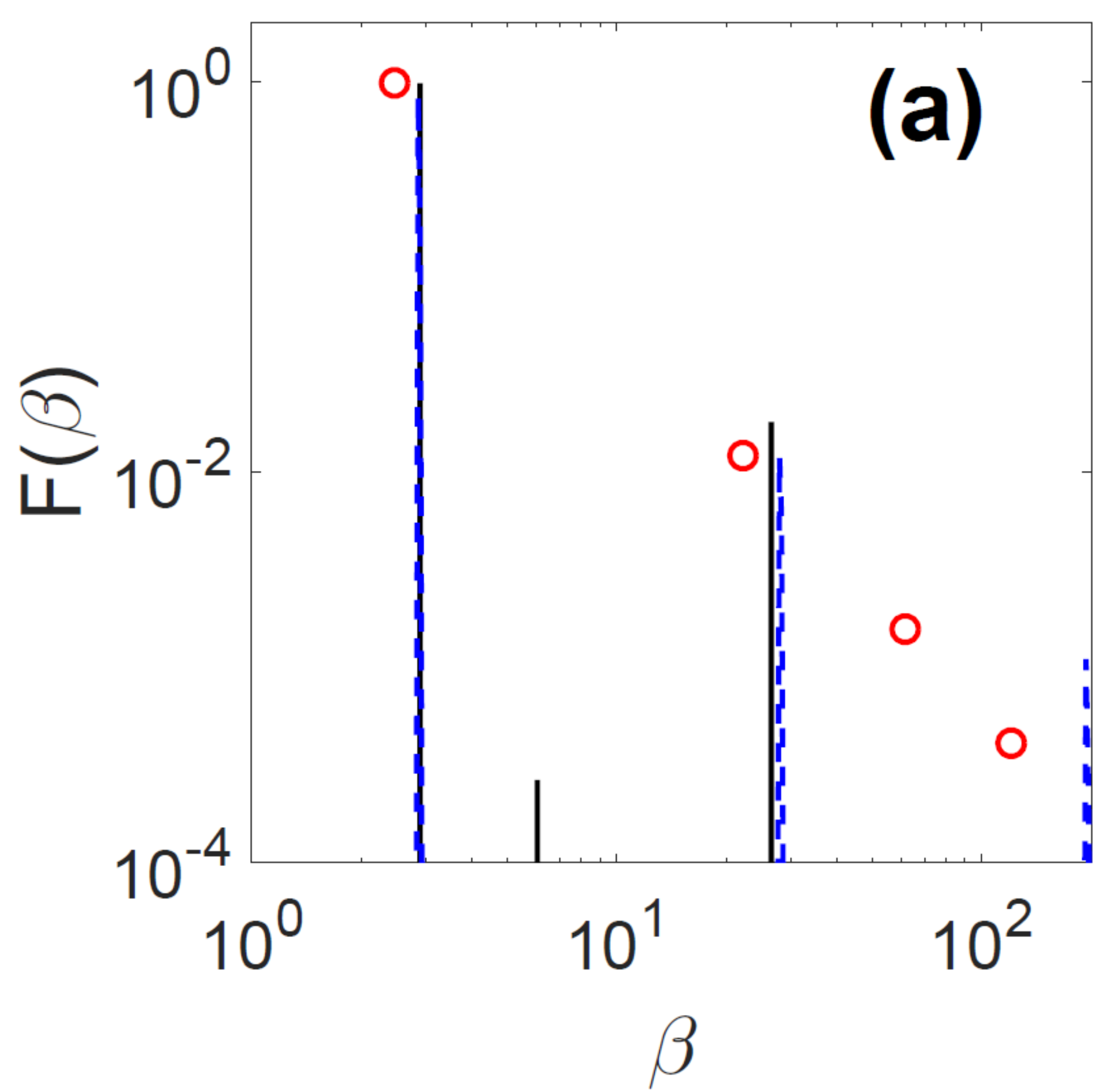}
}
\resizebox{0.23\textwidth}{!}{
\includegraphics{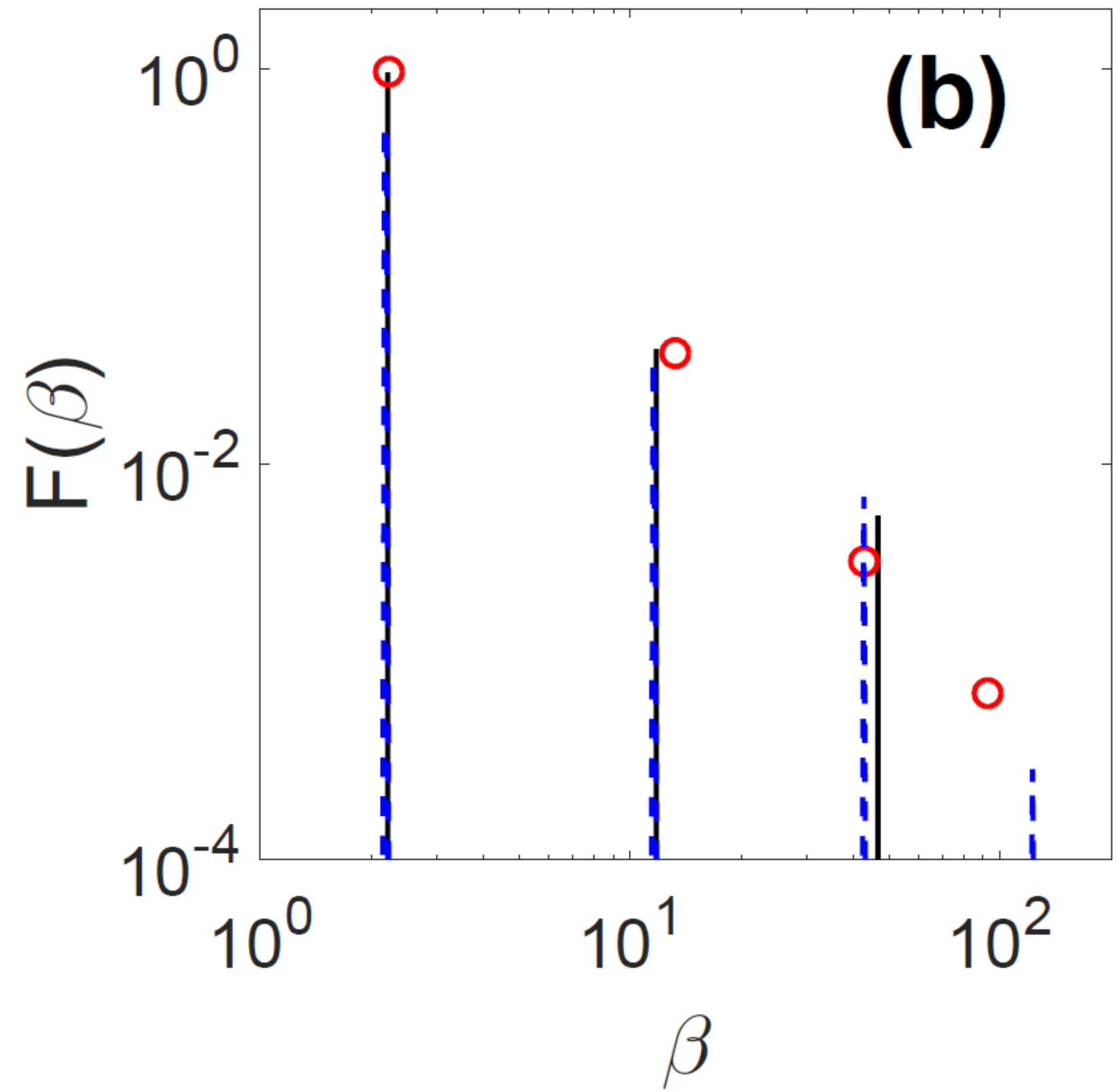}
}
%[clip, trim=0.5cm 11cm 0.5cm 11cm, width=1.00\textwidth]
%\includegraphics[scale=0.4]{T2_fig_8_a.pdf}~\hspace{-0mm}~\includegraphics[scale=0.4]{T2_fig_8_b.pdf}
%
\caption{Decomposition of the numerical total magnetisation of the ball (a)
and cube (b). Loglog plots of numerically obtained coefficients and
exponents, vertical (black) lines shows the discrete result from real
exponents ESPRIT, while the continuous (blue) dashed-curves are from
the Tikhonov regularized inversion. The four (red) circles in each
plot shows the four dominating analytic datapoints for the corresponding
geometries (from Table~\ref{Table_for_ball_and_cube}). Since the
LLBC is not exact for the ball, a small systematic mismatch in (a)
is expected for all terms. \label{fig:decomposition_of_cube_and_ball}}
\end{figure}

In Figure~\ref{fig:decomposition_of_analytic_ball} we compare the
analytic solution for the ball to the ESPRIT algorithm and the TRI
method. Since the analytic solution is used, ESPRIT is guaranteed
to find all exponents and coefficients within machine precision. In
Figure~\ref{fig:decomposition_of_cube_and_ball} we apply the real
exponents ESPRIT method and the TRI method to numerical data. The
numerical data is obtained using the methods introduced in Sect.~\ref{sec:Method}
to the ball (a) and the cube (b). The random walk method with LLBC
introduce small structural errors in the signal. This can be seen
as noise and a systematic deviation of the mean, respectively. Therefore,
it is not clear that analytic positions of the exponents give the
best fit to the numerical signal, i.e., that corresponding values
of $T_{j}$ and $d_{j}$ minimize~(\ref{eq:L2-norm}). Further testing
would be needed in order to draw conclusions about the reliability
of the two methods presented. The presented values also depend on
choosing certain parameters for both methods, however, Figure~\ref{fig:decomposition_of_cube_and_ball}
indicate that both methods do a good job in retrieving the two main
exponential terms in the signal.

\begin{table}
\tiny  
\begin{tabular}{|c|c|c|c|c|c|}
\hline 
Norm  & $\mathcal{N}_{ball,\, ana.}$  & $\mathcal{N}_{ball}$  & $\mathcal{N}_{cube}$  & $\mathcal{N}_{L}$  & $\mathcal{N}_{Aa}$\tabularnewline
\hline 
\hline 
$m=1$  & $1.3\cdot10^{-3}$  & $2.0\cdot10^{-3}$  & $4.8\cdot10^{-3}$  & $2.9\cdot10^{-4}$  & $3.2\cdot10^{-5}$ \tabularnewline
\hline 
$m=2$  & $3.6\cdot10^{-5}$  & $1.4\cdot10^{-4}$  & $3.5\cdot10^{-4}$  & $4.8\cdot10^{-5}$  & $3.1\cdot10^{-5}$ \tabularnewline
\hline 
$m=3$  & $9.0\cdot10^{-7}$  & $9.3\cdot10^{-5}$  & $1.6\cdot10^{-4}$  & $4.8\cdot10^{-5}$  & -\tabularnewline
\hline 
$m=4$  & $2.2\cdot10^{-14}$  & $9.0\cdot10^{-5}$  & $1.5\cdot10^{-4}$  & -  & -\tabularnewline
\hline 
$t_{2}$  & $0.927$  & $0.789$  & $1.02$  & $0.113$ {[}s{]}  & $0.0382$ {[}s{]} \tabularnewline
\hline 
\end{tabular}
\caption{\textit{Norm of the deviation between numerical data and the real exponents
ESPRIT approximation with $m$-terms, see Eq.~(\ref{eq:L2-norm}).
The first result column, $\mathcal{N}_{ball,\, ana.}$, reports the
norms when comparing an analytic series for the ball with four terms
(see Eq.~(\ref{eq:analytic_magnetization_for_ball}) and Table~\ref{Table_for_ball_and_cube})
with the few terms expansion~(\ref{eq:few_terms_representation})
according to ESPRIT. If we instead would have $m$-terms in the analytic
series for the ball, the ESPRIT method for $m$-terms is exact for
noiseless data up to numerical errors~\cite{Roy1989}, see row with
$m=4$ and the two subplots in Figure~\ref{fig:decomposition_of_analytic_ball}~(b).
The four remaining columns, $\mathcal{N}_{ball}$, $\mathcal{N}_{cube}$,
$\mathcal{N}_{L}$, and $\mathcal{N}_{Aa}$, reports the norms when
comparing the real exponents ESPRIT obtained approximation with $m$-terms
to numerical data for the ball, cube, limestone (L), and Aalborg chalk
(Aa), using LLBC in all four cases. For the chalk samples, which are
in the fast diffusing regime, the norm do not change substantially
by adding more than 2-3 terms (the two most right columns). The lengths
of all the vectors tested here are odd numbers in the order of $10^{3}$.
The truncation in the numerical data is set to the smallest time,
$t_{2}$, for which the total magnetisation fulfills $\mathcal{M}\left(t\right)/\mathcal{M}\left(0\right)<0.1$,
and is reported in the lowest row.} }
\label{Table_for_L2-norm} 
\end{table}

To further validate the result from the real exponents ESPRIT method
we calculate $\mathcal{N}$, for the corresponding few term approximations~(\ref{eq:few_terms_representation}),
according to the discrete version of~(\ref{eq:L2-norm}) for different
number of terms, $m$, and with a relatively large upper integration
limit, $t_{2}$, specified in Table~\ref{Table_for_L2-norm}. Since
the result of the real exponents ESPRIT algorithm is dependent on
a parameter $\tau$ and the number of iterations, we stress that the
reported norms are not unique. The lowest exponents obtained with
real exponents ESPRIT for the chalk samples, $\beta_{1}\simeq20.4$~s$^{-1}$
for Limestone and $\beta_{1}\simeq60.3$~s$^{-1}$ for Aalborg chalk,
is in agreement with the estimates of the half-times $T_{1/2}=\ln2/\beta_{1}$
given in Sect.~\ref{sub:Results-of-the-total-magnetization-in-chalk}.

\section{Discussion and summary\label{sec:Discussion-and-summary}}

For the NMR relaxation simulation described here, we started from
a deterministic partial differential equation and then used an equivalent
stochastic particle formulation for the calculation. As expected from
the previous investigation~\cite{Ogren2014}, the systematic errors
for the NMR relaxation caused by the digitalisation of the 2D surfaces
within the 3D geometrical objects occurs qualitatively different depending
on the object and its orientation in relation to the coordinate system
in a digital image. Here we have quantified those systematic errors
and showed how they can be reduced for the ball and the cube in 3D
and for an artificial 2D porous media for which comparative finite
element calculations were tractable.

For two complex digital domains representing different chalk samples,
that had similar porosity, but with substantially different specific
area, we found qualitatively different relaxation dynamics. Additionally
for each of those complex domains the relaxation curve without local
boundary conditions were markably lower and we expect to have removed
a major part of the errors between the true NMR relaxation and its
simulated dynamics.

For the inversion analyse of the NMR relaxation data, commonly called
$T$-distribution curve, we have introduced and benchmarked a new
method to the field.

\section*{Acknowledgement}

The authors from the University of Copenhagen would like to thank
Innovation Fund Denmark and Maersk Oil for funding this research through
the project P$^{3}$ \textemdash{} \emph{Predicting Petrophysical
Parameters.} We acknowledge Danscatt for supporting datasampling and
thank Ye Zhang for providing software for the Tikhonov regularization.

%% References


%%\bibliographystyle{elsarticle-num}
%%\bibliography{<your-bib-database>}



\end{document}